\documentclass[preprint,aps,amsmath,showpacs,floatfix]{revtex4}
\usepackage{graphicx}
\usepackage[latin1]{inputenc}
\begin{document}

\title{Quantum Dissipation and Decoherence via Interaction with
Low-Dimensional Chaos: a Feynman-Vernon Approach}

\author{M.V.S. Bonan\c{c}a and M.A.M. de Aguiar}

\affiliation{Instituto de F\'isica 'Gleb Wataghin',
Universidade Estadual de Campinas, \\
Caixa Postal 6165, 13083-970 Campinas, S\~ao Paulo, Brazil}

\begin{abstract}

We study the effects of dissipation and decoherence induced on a
harmonic oscillator by the coupling to a chaotic system with two
degrees of freedom. Using the Feynman-Vernon approach and treating
the chaotic system semiclassically we show that the effects of the
low dimensional chaotic environment are in many ways similar to
those produced by thermal baths. The classical correlation and
response functions play important roles in both classical and
quantum formulations. Our results are qualitatively similar to the
high temperature regime of the Caldeira-Leggett model.

\end{abstract}

\maketitle

%%%%%%%%%%%%%%%%%%%%%%%%%%%%%%%%%%%%%%%%%%%%%%%%%%%%%%%%%%%%%%
%%%%%%%%%%%%%%%%%%%%%%%%%%%%%%%%%%%%%%%%%%%%%%%%%%%%%%%%%%%%%%
\section{Introduction.}

The relation between chaos and the phenomena of quantum dissipation
and decoherence has attracted a lot of attention in the last ten
years
\cite{cohen1,cohen2,cohen3,wilkinson,berry1,jarzynski,tulio,zurek1,zurek2,zurek3,cohen4,paz}.
The problem considered in most works involves the weak interaction
of a chaotic system with an external oscillator. Various points of
view have been considered by different authors. One approach
concentrates on the chaotic system itself, focusing on the
dissipation and treating the external oscillator as a time dependent
parameter that perturbs the chaotic system
\cite{cohen1,cohen2,cohen3}. The basic assumption is that the
external system is slow and sufficiently heavy not to be affected
significantly by the coupling
\cite{wilkinson,berry1,jarzynski,tulio}. The effect on the chaotic
system, on the other hand, is that of an adiabatic perturbation.
Under these conditions, a dissipative force, acting on the external
system, may result.

A different view of the same problem focus on the semiclassical
limit of chaotic systems. It has been shown
\cite{zurek1,zurek2,zurek3,cohen4,paz} that the coupling of a
chaotic system with an external environment, represented implicitly
by a small diffusion constant in the classical and quantum versions
of the Focker-Planck equation, leads to a very close correspondence
between the classical and quantum evolutions. The coupling causes
destruction of quantum interference and, at the same time, it washes
out the fine structures of the classical distributions, bringing the
two dynamics together.

In this paper we consider the chaotic system and the external
oscillator explicitly, as a single globally conservative Hamiltonian
system. We use the Feynman-Vernon approach to trace out the chaotic
system variables and construct an effective dynamics for the
oscillator, in close analogy with the treatment of the Brownian
motion considered by Caldeira and Leggett
\cite{caldeira1,caldeira2}. While we focus on the oscillator,
looking at dissipation and decoherence, the effects of the coupling
on the chaotic system are also taken into account consistently. This
characteristic of the Feynman-Vernon method turns out to be very
important in this problem, since both the oscillator and the chaotic
system have small number of degrees of freedom and are both affected
by the mutual interaction. However, whereas the treatment of
Caldeira and Leggett is phenomenological, in the sense that the
spectral properties of the reservoir are not derived from its
Hamiltonian, the case of a small chaotic environment has to be
treated dynamically. In other words, dissipation and decoherence
have to come out directly from correlations and response functions.

Our purpose here is to understand under what conditions a chaotic
system with only two degrees of freedom can produce dissipation and
decoherence, phenomena usually related to many body thermal baths.
In a previous paper \cite{bonanca} we have considered the
interaction of an oscillator with a chaotic system from a classical
point of view. We showed that the effects of the oscillator on the
environment cannot be neglected. Here we consider the quantum
version of the same problem, assuming the chaotic system to be in
the semiclassical regime.

In treatment of the Brownian motion by Caldeira and Leggett, the
degrees of freedom playing the role of the environment are averaged
with a canonical ensemble, since the reservoir is kept at the
constant temperature. Here, since the environment is small, the
microcanonical ensemble is more adequate. A similar approach was
recently considered by Esposito and Gaspard using random matrix
theory to model the chaotic environment
\cite{esposito1,esposito2,esposito3}.

The paper is organized as follows: in Sec.II, we review some aspects
of the classical formulation that are useful for the quantum
analysis. In Sec.III, we present the quantum formulation using
Feynman-Vernon approach \cite{feynman1}. The formal development
leads to quantum correlation and response functions, that we
calculate semiclassically. In Sec.IV, we consider two basic
applications: first, the propagation of an Gaussian state, where we
characterize quantum dissipation. Second, we calculate the evolution
of a superposition of two Gaussian states, focusing on the
decoherence due to the chaotic environment. In section V we present
our conclusions.

%%%%%%%%%%%%%%%%%%%%%%%%%%%%%%%%%%%%%%%%%%%%%%%%%%%%%%%%%%%%%%
%%%%%%%%%%%%%%%%%%%%%%%%%%%%%%%%%%%%%%%%%%%%%%%%%%%%%%%%%%%%%%
\section{Classical Formulation}

In this section we describe the behavior of a system of interest
coupled to a small chaotic environment from the classical point of
view. Although the formalism presented here can be extended to more
general systems, we particularize right away to the case of a
harmonic oscillator interacting with the so called Nelson system.
The discussion outlined here is a summary of the detailed results
presented in ref.\cite{bonanca} (see also
\cite{wilkinson,berry1,jarzynski,tulio}). The Hamiltonian of the
system is given by
\begin{eqnarray}
H = H_o(z) + H_c(x,y) + V_I(x,z), \label{eq1}
\end{eqnarray}
where
\begin{equation}
H_o(z) = \frac{p_z^2}{2m} + \frac{m\omega_0^2 z^2}{2},
\end{equation}
represents the system of interest,
\begin{equation}
V_I(x,z) = \gamma xz,
\end{equation}
is the interaction potential and
\begin{equation}
H_c(x,y) = \frac{p_x^2}{2} + \frac{p_y^2}{2} + \left(y - \frac{x^2}
{2}\right)^2 + 0.1\frac{x^2}{2}
\end{equation}
is the chaotic Hamiltonian, known as Nelson system (NS)
\cite{baranger}. The NS exhibits soft chaos and is fairly regular
for $E_c \,_{\sim}^<\,0.05$, strongly chaotic for $E_c \,_{\sim}^>\,
0.3$ and mixed for intermediate values of the energy.

In order to investigate the situation where $H_c$ plays the role of
an external environment for the oscillator, we assume that detailed
information about the chaotic system is not available. If the
environment were modeled by a heat bath, the only macroscopic
relevant information would be its temperature. In the present case
we assume that the only information available is the initial energy
$E_c(0)$ of the chaotic system. For the oscillator this implies that
only averages of its observables (over the chaotic system variables)
are accessible.

When the coupling between the chaotic system and the oscillator is
turned on, the overall conserved energy flows from one system to the
other. The oscillator's energy, in particular, fluctuates as a
function of the time for each specific trajectory. The oscillator's
average energy is calculated by taking an ensemble of initial
conditions uniformly distributed over the chaotic energy surface
$E_c(0)$. For the oscillator we fix only one initial condition,
which we choose to be $z(0) = 0$ and $p_z(0) = \sqrt{2mE_o (0)}$.
The microcanonical ensemble of chaotic initial conditions plus the
fixed oscillator's initial condition are propagated numerically and,
at each instant, $H_o$ is calculated for each trajectory and its
average value is computed. We have shown in \cite{bonanca} that the
oscillator's average energy $\langle E_o (t)\rangle$ tends to a
constant value for long times, indicating a 'thermalization' of the
systems.

The short time behavior of $\langle E_o(t)\rangle$ can be obtained
from the Linear Response Theory \cite{kubo}. From the equations of
motion for $z$ and $p_z$ we find
\begin{eqnarray}
z(t)=z_d (t)-\frac{\gamma}{m}\int_0^t
\mathrm{d}s\Gamma(t-s)x(s),\label{eq3} \\
p_z (t) = p_{z_d} (t)-\gamma\int_0^t \mathrm{d}s\chi(t-s)x(s), \label{eq4}
\end{eqnarray}
where $z_d (t)$ and $p_{z_d} (t)$ are the decoupled solutions, given
by,
\begin{equation}
z_d (t)=\frac {p_z (0)}{m\omega_0}\sin{(\omega_0t)} \qquad p_{z_d}
(t)=p_z (0)\cos{(\omega_0t)}\;,
\end{equation}
where $\Gamma(t-s) = \sin{[\omega_0(t-s)]}/\omega_0$ and $\chi(t-s)
= \cos{[\omega_0 (t-s)]}$.
Thus,
\begin{eqnarray}
\label{eq5}
\langle p_z^2 (t)\rangle = p^2_{z_d}(t)-2\gamma
p_{z_d} (t)\int_0^t
\mathrm{d}s\chi(t-s)\langle x(s)\rangle+\nonumber \\
\gamma^2\int_0^t\mathrm{d}s\int_0^t\mathrm{d}u
\chi(t-s)\chi(t-u)\langle x(s)x(u)\rangle
\end{eqnarray}
and
\begin{eqnarray}
\label{eq6} \langle z^2 (t)\rangle=z^2_{d}(t)-\frac{2\gamma}{m}
z_d (t)\int_0^t
\mathrm{d}s\Gamma(t-s)\langle x(s)\rangle+\nonumber \\
\frac{\gamma^2}{m^2}\int_0^t\mathrm{d}s\int_0^t\mathrm{d}u \Gamma(t-s)
\Gamma(t-u)\langle x(s)x(u)\rangle.
\end{eqnarray}
The oscillator's average energy can now be obtained from
\begin{equation}
\langle E_o (t)\rangle=\frac{\langle p_z^2 (t)\rangle}{2m}+\frac{m
\omega_0^2 \langle z^2 (t)\rangle}{2} \label{eq2}
\end{equation}

Equations (\ref{eq5}) and (\ref{eq6}) show that we need $\langle
x(t)\rangle$ and $\langle x(0)x(t)\rangle$ in order to calculate
$\langle E_o(t)\rangle$. The calculation of such averages involve
the distribution function $\rho(q,p;t)$ whose initial value is
$\rho(q,p;0) = \delta(H_c(q,p)-E_c(0))/\Sigma(E_c(0))$, with
$\Sigma(E_c(0))=\int \mathrm{d}q\mathrm{d}p\, \delta
(H_c(q,p)-E_c(0))$. Here we are using $q=(x,y)$ and $p=(p_x,p_y)$
for the coordinates and momenta of the chaotic system. If the
chaotic system were isolated, $\rho$ would be an invariant
distribution and $\rho(q,p;t)=\rho(q,p;0)$. The coupling, however,
causes the value of $H_c(q(t),p(t))$ to fluctuate in time,
distorting the energy surface $H_c=E_c(0)$. Linear Response Theory
provides the first order corrections to this distribution in the
limit of weak coupling \cite{kubo}. Keeping in (\ref{eq5}) and
(\ref{eq6}) only terms up to order $\gamma^2$, we find
\begin{eqnarray}
\langle x(t)\rangle=\langle x(t)\rangle_e -
\gamma\int_0^t\mathrm{d}s \phi_{xx} (t-s)z(s) \label{eq7}
\end{eqnarray}
and
\begin{eqnarray}
\langle x(0)x(t)\rangle=\langle x(0)x(t)\rangle_e, \label{eq8}
\end{eqnarray}
where $\langle A(q,p)\rangle_e=\int\mathrm{d}q\mathrm{d}p\,A(q,p)
\rho_e(q,p)$ and $\rho_e=\delta(H_c-E_c(0))/\Sigma(E_c(0))$ is the
microcanonical distribution of the isolated chaotic system.
$\phi_{xx} (t)$ is the response function, given by \cite{kubo}
\begin{eqnarray}
\phi_{xx}(t)=\langle \{x(0),x(t)\}\rangle_e=\int\int\mathrm{d}V\,
\rho_e\,\{x(0),x(t)\} \label{eq9}
\end{eqnarray}
where $\mathrm{d}V = \mathrm{d}x(0) \mathrm{d}y(0) \mathrm{d}
p_x(0)\mathrm{d}p_y(0)$ and $\{.,.\}$ is the Poisson bracket with
respect to the initial conditions. Since $H_c(-x)=H_c(x)$, $\langle
x(t)\rangle_e=0$. Substituting (\ref{eq7}) and (\ref{eq8}) into
(\ref{eq5}) and (\ref{eq6}) we obtain
\begin{equation}
\begin{array}{ll}
\langle p^2_z (t)\rangle &=
p^2_{z_d}(t)+2\gamma^2 p_{z_d}(t)\int_0^t\mathrm{d}s\chi(t-s)
\int_0^s\mathrm{d}u\phi_{xx}(s-u)z_d(u)  \\
 & + \gamma^2\int_0^t \mathrm{d}s\int_0^t
\mathrm{d}u\chi(t-s)\chi(t-u) \langle x(s)x(u)\rangle_e,
\label{eq10}
\end{array}
\end{equation}
\begin{equation}
\begin{array}{ll}
\langle z^2(t)\rangle & =
z^2_d(t)+\frac{2\gamma^2}{m}z_d(t)\int_0^t\mathrm{d}s
    \Gamma(t-s)\int_0^s\mathrm{d}u\phi_{xx}(s-u)z_d(u) \\
&+\frac{\gamma^2}{m^2}\int_0^t \mathrm{d}s \int_0^t
  \mathrm{d}u\Gamma(t-s)\Gamma(t-u)\langle x(s)x(u) \rangle_e.
\label{eq11}
\end{array}
\end{equation}

Eqs. (\ref{eq10}) and (\ref{eq11}) show that all the influence of
the chaotic system is contained in the functions $\langle
x(0)x(t)\rangle_e$ and $\phi_{xx}(t)$. For NS, the response function
is given by \cite{bonanca}
\begin{equation}
\phi_{xx}(t)=\frac{2}{E_c(0)}\langle p_x(0)x(t)\rangle_e.
\label{eq-resp}
\end{equation}
The correlation functions $\langle p_x(0)x(t)\rangle_e$ and $\langle
x(0)x(t)\rangle_e$ are obtained numerically with a fixed time step
symplectic integration algorithm \cite{forest} applied to the
isolated chaotic system. Fig. 1 shows the numerical correlation
functions for $E_c(0)=0.38$. These numerical results can be well
fitted by the expressions
\begin{equation}
\begin{array}{ll}
\langle x(0)x(t)\rangle_e &=\sigma e^{-\alpha t} \cos{\omega t},\\
\langle p_x (0) x(t)\rangle_e &=\mu e^{-\beta t} \sin{\Omega t},
\label{eq12}
\end{array}
\end{equation}
with decay rates $\alpha=0.0418$ and $\beta=0.0456$, amplitudes
$\sigma=1.865$ and $\mu=0.409$ and frequencies of oscillation
$\omega=0.1963$ and $\Omega=0.2043$ with $\chi^2\,\sim\,10^{-4}$
(see Fig.1). Notice that the exponents $\alpha$ and $\beta$ and the
frequencies $\omega$ and $\Omega$ are very similar. If the
interacting system were integrable, these functions would exhibit
quasi-periodic oscillations.

Considering the expressions (\ref{eq12}) and assuming
$\Omega\approx\omega$ and $\beta\approx\alpha$, we obtain the
following result for $\langle E_o(t)\rangle$
\begin{eqnarray}
\langle E_o(t)\rangle = E_o(0)+\frac{\gamma^2}{m}(B+At+f(t)+g(t)),
\label{eq13}
\end{eqnarray}
where $B$ is a constant, $f(t)$ is an oscillatory function and
$g(t)$ is proportional to $e^{-\alpha t}$. The important result is
the coefficient $A$
\begin{equation}
A=4\mu\omega\alpha \frac{\left[\frac{\sigma}{4\mu\omega}
(\omega^2_0+\omega^2+\alpha^2)-\frac{E_o(0)}
{E_c(0)}\right]}{[(\omega_0-\omega)^2+\alpha^2][(\omega_0+
\omega)^2+\alpha^2]}. \label{eq14}
\end{equation}
For fixed oscillator frequency $\omega_0$ and a given chaotic
energy shell $E_c(0)$ (and, consequently, for given $\sigma$,
$\mu$, $\omega$ and $\alpha$), the ratio $E_o(0)/E_c(0)$ is the
responsible for the average increase or decrease of $\langle
E_o(t)\rangle$. The short time equilibrium in the energy flow is
given by the condition $A=0$, or
\begin{equation}
\frac{E_o(0)}{E_c(0)}=\frac{\sigma}{4\mu\omega}(\omega^2_0+\omega^2
+\alpha^2) \;. \label{eq15}
\end{equation}
The equation of motion of $z(t)$ under the average effect of the
interaction with the chaotic system can also be written in terms of
the response function as
\begin{equation}
\langle\ddot{z}(t)\rangle+\omega^2_0 \langle z(t)\rangle
=-\frac{\gamma}{m}\langle x(t)\rangle
  =\frac{\gamma^2}{m}\int_0^t\mathrm{d}s \, \phi_{xx} (t-s)
  \langle z(s)\rangle.
\end{equation}
Integrating by parts yields
\begin{eqnarray}
\langle\ddot{z}(t)\rangle+\left(\omega^2_0-\frac{\gamma^2
F(0)}{m}\right) \langle z(t)\rangle+
\frac{\gamma^2}{m}\int_0^t\mathrm{d}s \, F(t-s) \langle
 \dot{z}(s)\rangle+ \frac{\gamma^2}{m}z(0)F(t)=0, \label{eq16}
\end{eqnarray}
where
\begin{equation}
F(t-s) =\int\mathrm{d}s\,\phi_{xx}(t-s)=\frac{2\mu
e^{-\alpha(t-s)}}{E_c(0)(\alpha^2+\omega^2)}\{\omega
\cos{[\omega(t-s)]} +\alpha\sin{[\omega(t-s)]}\}. \label{eq17}
\end{equation}

Eq.(\ref{eq16}) shows that the interaction produces a harmonic
correction to the original potential, a dissipative term with memory
and an external force proportional to $z(0)$. The choice $z(0)=0$
simplifies (\ref{eq16}) and turns it into an average Langevin
equation.

Fig.2 shows a comparison between the numerically calculated `bare'
oscillator energy $\langle E_o(t)\rangle$, where
\begin{equation}
E_o(z) = \frac{p_z^2}{2m} + \frac{m\omega_0^2 z^2}{2},
\end{equation}
the `re-normalized' oscillator energy $\langle E_{or}(t)\rangle$,
where
\begin{equation}
E_{or}(z) = \frac{p_z^2}{2m} + \frac{m\omega_0^2 z^2}{2}
-\frac{\gamma^2}{2}F(0)z^2, \label{eq18}
\end{equation}
and the expression (\ref{eq13}) without the oscillating term $f(t)$.
We have chosen $\gamma$ and $m$ so that $\omega^2_0 - \gamma^2
F(0)/m>0$. We also have chosen $\omega_0$ so that
$e^{-\alpha/\omega_0}\approx 10^{-4}$ and $g(t)$ decreases very
fast. In this case only the linear and the oscillating terms in
Eq.(\ref{eq13}) are important. We have subtracted the oscillating
part of Eq.(\ref{eq13}) in Fig.2 to highlight the linear increase or
decrease in the average energy. In the time scale of Fig.2, which
corresponds to several periods of the decoupled oscillator, the
linear behavior describes very well the numerical results. Fig.2(b)
shows the equilibrium situation according to Eq.(\ref{eq15}). Notice
that $F(t)$ decays very fast in the time scale $1/\omega_0$, leading
to the dissipative force in Eq.(\ref{eq16}).

In the next sections we consider the quantum counterpart of these
classical calculations. The chaotic system will be treated
semiclassically and the quantum versions of the response and
correlation functions will play important roles.

%%%%%%%%%%%%%%%%%%%%%%%%%%%%%%%%%%%%%%%%%%%%%%%%%%%%%%%%%%%%%%
%%%%%%%%%%%%%%%%%%%%%%%%%%%%%%%%%%%%%%%%%%%%%%%%%%%%%%%%%%%%%%
\section{Quantum Formulation}

%%%%%%%%%%%%%%%%%%%%%%%%%%%%%%%%%%%%%%%%%%%%%%%%%%%%%%%%%%%%%%
\subsection{The Feynman-Vernon Approach}

In this section we describe the dynamics of the coupled oscillator
from a quantum point of view. In order to do that we need, like in
the classical case, a systematic way to eliminate the detailed
information we don't need about the chaotic system. We will do that
using the Feynman-Vernon approach \cite{feynman1}. Because of the
non-linear chaotic system, we will not be able to perform exact
calculations. Instead, we will resort to semiclassical
approximations.

We consider the quantum version of the full Hamiltonian,
Eq.(\ref{eq1}), and again we denote by $q=(x,y)$ the pair of
coordinates of the chaotic system. The density matrix operator can
be written as
\begin{eqnarray}
\hat{\rho}(T)=|\psi(T)\rangle \langle\psi(T)| = e^{-i\hat{H}T/\hbar}
|\psi(0)\rangle \langle\psi(0)|e^{i\hat{H}T/\hbar} \nonumber
\end{eqnarray}
where $\psi(T)$ is the wave function of the whole system. In the
position representation
\begin{eqnarray}
\lefteqn{\rho(z(T),q(T),z'(T),q'(T))=\langle z,q|\psi(T)\rangle
\langle\psi(T)|z',q'\rangle} \nonumber \\
&=&\int\mathrm{d}z(0)\mathrm{d}z'(0)\mathrm{d}q(0)\mathrm{d}q'(0)
\langle z,q|e^{-i\hat{H}T/\hbar}|z(0),q(0)\rangle\langle z(0),q(0)|
\psi(0)\rangle \times \nonumber \\
& &\langle\psi(0)|z'(0),q'(0)\rangle\langle z'(0),q'(0)|
e^{i\hat{H}T/\hbar}|z',q'\rangle \nonumber \\
&=&\int\mathrm{d}z(0)\mathrm{d}z'(0)\mathrm{d}q(0)\mathrm{d}q'(0)
K(z(T),q(T),z(0),q(0))\psi(z(0),q(0)) \times \nonumber \\
& &K^*(z'(T),q'(T),z'(0),q'(0))\psi^*(z'(0),q'(0)),
\end{eqnarray}
where the propagators can be written in terms of Feynman path
integrals as \cite{feynman2}
\begin{eqnarray}
K(z(T),q(T),z(0),q(0))=\int\mathrm{D}z(t)\mathrm{D}q(t)\exp{\left[
\frac{i}{\hbar}S[z(t),q(t)]\right]},
\end{eqnarray}
with action
\begin{eqnarray}
S[z(t),q(t)]&=& \int_0^T\mathrm{d}t
(L_0(z(t))+L_c(q(t))+L_I(z(t),q(t))) \nonumber \\
&\equiv &S_o+S_c+S_I.
\end{eqnarray}
Thus,
\begin{eqnarray}
\lefteqn{\rho(z(T),q(T),z'(T),q'(T))=} \nonumber \\
&=&\int\mathrm{d}z(0)\mathrm{d}z'(0)
\mathrm{d}q(0)\mathrm{d}q'(0)\mathrm{D}z(t)\mathrm{D}z'(t)\mathrm{D}q(t)
\mathrm{D}q'(t)\nonumber \\
&
&\exp\left[\frac{i}{\hbar}\left(S[z(t),q(t)]-S[z'(t),q'(t)]\right)\right]
\rho(z(0),q(0),z'(0),q'(0)),
\end{eqnarray}
where
\begin{equation}
\rho(z(0),q(0),z'(0),q'(0))=\psi(z(0),q(0))\psi^*(z'(0),q'(0))
\end{equation}
is the initial state. As usual we use $z(T)$ and $q(T)$ just as
labels for the positions $z$ and $q$ at $T$. $z(T)$ and $q(T)$ are
not functions of $T$.

We assume that the initial state can be written as
\begin{equation}
\rho(z(0),q(0),z'(0),q'(0))=\rho_o(z(0),z'(0))\rho_c(q(0),q'(0))
\end{equation}
and define the reduced density matrix by
\begin{equation}
\rho_o(z(T),z'(T))=\int\mathrm{d}q(T)\rho(z(T),q(T),z'(T),q(T)).
\end{equation}
We obtain
\begin{eqnarray}
\lefteqn{\rho_o(z(T),z'(T))=\int\mathrm{d}z(0)\mathrm{d}z'(0)\mathrm{D}z(t)
\mathrm{D}z'(t)}  \nonumber \\
& &\left\{\int\mathrm{d}q(0)\mathrm{d}q'(0)\mathrm{d} q(T)
\mathrm{d}q'(T)\mathrm{D}q(t)
\mathrm{D}q'(t)\delta(q(T)-q'(T))\right.
\nonumber \\
& &\exp{\left[\frac{i}{\hbar}(S_c[q(t)]-S_c[q'(t)]+S_I[z(t),
q(t)] -S_I[z'(t),q'(t)\right]} \nonumber\\
& &\rho_c(q(0),q'(0)) \bigg\} \exp{\left[\frac{i}
{\hbar}\left(S_o[z(t)]-S_o[z'(t)]\right)\right]\rho_o(z(0),z'(0))},
\end{eqnarray}
which can be written as
\begin{equation}
\rho_o(z(T),z'(T))=\int\mathrm{d}z(0)\mathrm{d}z'(0)J(z(T),z'(T),z(0),
z'(0))\rho_o(z(0),z'(0)), \label{eq19}
\end{equation}
where
\begin{equation}
J(z(T),z'(T),z(0),z'(0))=\int\mathrm{D}z(t)\mathrm{D}z'(t)\mathcal{F}
[z(t),z'(t)]\exp[\frac{i}{\hbar}(S_o[z(t)]-S_o[z'(t)])]\label{eq20}
\end{equation}
is the `superpropagator' and
\begin{eqnarray}
\lefteqn{\mathcal{F}[z(t),z'(t)]=\int\mathrm{d}q(0)\mathrm{d}q'(0)
\mathrm{d}q(T)
\mathrm{d}q'(T)\mathrm{D}q(t)\mathrm{D}q'(t)\delta(q(T)-q'(T))}
\nonumber \\ &
&\exp\left[\frac{i}{\hbar}\left(S_c[q(t)]-S_c[q'(t)]+S_I[z(t),q(t)]-
S_I[z'(t),q'(t)]\right)\right] \; \rho_c(q(0),q'(0)),
\end{eqnarray}
is the so called `influence functional', which contains all the
information about the chaotic system.

Equation (\ref{eq19}) is the equation of motion for the reduced
density matrix. For $\mathcal{F}=1$, $J$ becomes the propagator of
the isolated oscillator. Our goal is to get an approximate
expression for $J$ that includes the effects of the chaotic system.

We take as the initial state for the chaotic system one of its
energy eigenstate $\phi_a(q)=\langle q|a\rangle$. Thus,
\begin{equation}
\rho_c(q(0),q'(0))=\phi_a(q(0))\phi_a^*(q'(0)).
\end{equation}
This is the quantum version of the classical microcanonical
distribution we considered in section II.

The difficulty in the calculation of $J$ is that the chaotic
Lagrangian $L_c$ is not quadratic. Therefore, $\mathcal{F}$ has to
be treated in a perturbative manner. Rewriting $\mathcal{F}$ as
\begin{eqnarray}
\lefteqn{\mathcal{F}[z(t),z'(t)]=\int\mathrm{d}q(0)\ldots
\mathrm{D}q'(t)\delta(q(T)-
q'(T))\exp\left[\frac{i}{\hbar}(S_c[q(t)]-S_c[q'(t)])\right]}\nonumber\\
&
&\exp\left[-\gamma\frac{i}{\hbar}\left(\int_0^T\mathrm{d}t(z(t)x(t)-
z'(t)x'(t))\right)\right]\phi_a(q(0))\phi_a^*(q'(0))
\end{eqnarray}
we assume that $\gamma$ is small enough so that the exponential in
the second line can be expanded to second order in its argument.
These terms can be calculated by inserting complete sets of energy
eigenstates of $H_c$. The result, following \cite{feynman1}, is
\begin{eqnarray}
\lefteqn{\mathcal{F}[z(t),z'(t)] \approx
1-\left(\frac{i\gamma}{\hbar}\right)x_{aa}
\int_0^T\mathrm{d}t[z(t)-z'(t)]} \nonumber \\
& &-\left(\frac{\gamma^2}{\hbar}\right)
\int_0^T\mathrm{d}t\int_0^t\mathrm{d}s[z(t)-z'(t)][z(s)F_a^*(t-s)
-z'(s)F_a(t-s)],
\end{eqnarray}
where
\begin{eqnarray}
F_a(t-s) = \sum_b\frac{|x_{ba}|^2}{\hbar}\exp[i\omega_{ba}(t-s)],
\qquad\omega_{ba}=\frac{E_b-E_a}{\hbar},\label{eq21}
\end{eqnarray}
\begin{eqnarray}
x_{ba}=\int\mathrm{d}q\phi_b^*(q)x\phi_a(q),
\end{eqnarray}
and $E_b$ are the eigen-energies of chaotic system ($x$ is the
coordinate of $H_c$ in $V_I$ ). For the NS $H_c(-x)=H_c(x)$ and
$x_{aa}=0$. Thus,
\begin{eqnarray}
\mathcal{F}[z(t),z'(t)] \approx 1-\frac{1}{\hbar}\Phi[z(t),z'(t)]
\approx \exp\left[-\frac{1}{\hbar}\Phi[z(t),z'(t)]\right]
\end{eqnarray}
where
\begin{eqnarray}
\Phi[z,z'] = \frac{\gamma^2}{\hbar}
\int_0^T\mathrm{d}t\int_0^t\mathrm{d}s[z(t)-z'(t)][z(s)F_a^*(t-s)-
z'(s)F_a(t-s)].
\end{eqnarray}

With these approximations the superpropagator can be written as
\begin{eqnarray}
J(z(T),z'(T),z(0),z'(0))=\int\mathrm{D}z(t)\mathrm{D}z'(t) \, e^{
\frac{i}{\hbar}(\tilde{S}_{ef}[z(t),z'(t)])},
\end{eqnarray}
where we have defined the effective action
\begin{equation}
\tilde{S}_{ef}[z(t),z'(t)]=S_o[z(t)]-S_o[z'(t)]+i\Phi[z(t),z'(t)].
\label{sefect}
\end{equation}

Since $\tilde{S}_{ef}$ is quadratic in $z$ and $z'$ the path
integral can be solved exactly by the stationary phase method. It is
convenient to define the new variables $r(t)=(z(t)+z'(t))/2$ and
$y(t)=z(t)-z'(t)$ \cite{weiss} and to separate
$F_a(t)=F_a^{'}(t)+iF_a^{''}(t)$ into real and imaginary parts. This
allows us to write
\begin{eqnarray}
J(r(T),y(T),r(0),y(0))= \int\mathrm{D}r(t)\mathrm{D}y(t)\, e^{
\frac{i}{\hbar}\tilde{S}[r(t),y(t)] \, - \, \frac{1}{\hbar}
\phi[r(t),y(t)]} \label{pathint}.
\end{eqnarray}
where
\begin{eqnarray}
\tilde{S}[r(t),y(t)]\equiv\int_0^T\mathrm{d}t\left\{m[\dot{r}(t)\dot{y}(t)
-\omega_0^2r(t)y(t)]+2\gamma^2y(t)\int_0^t\mathrm{d}sF_a^{''}(t-s)
r(s)\right\}, \label{sefreal}
\end{eqnarray}
and
\begin{eqnarray}
\phi[r(t),y(t)]\equiv \gamma^2\int_0^T\mathrm{d}t\int_0^t\mathrm{d}s
y(t)y(s)F_a^{'}(t-s), \label{sefimag}
\end{eqnarray}
are the real and imaginary parts of $\tilde{S}_{ef}$. In Appendix A
we show that
\begin{eqnarray}
J(r(T),y(T),r(0),y(0))=G(T,0)\exp\left\{\frac{i}{\hbar}\tilde{S}[r_e,y_e]
\right\}\exp\left\{-\frac{1}{\hbar}\phi[y_e,y_e]\right\},
\label{superprop}
\end{eqnarray}
where $G(T,0)$ can be obtained by the normalization condition of
reduced density matrix and $r_e(t)$ and $y_e(t)$ are the extremum
paths of $\tilde{S}$, which satisfy
\begin{eqnarray}
\ddot{r}_e(t)+
\omega_0^2r_e(t)-\frac{2\gamma^2}{m}\int_0^t\mathrm{d}sF_a{''}(t-s)
r_e(s)=0,\label{eq22} \\
\ddot{y}_e(t)+
\omega_0^2y_e(t)-\frac{2\gamma^2}{m}\int_0^t\mathrm{d}sF_a{''}(s-t)
y_e(s)=0.\label{eq23}
\end{eqnarray}
Therefore we need $F_a^{''}$ to solve (\ref{eq22}) and (\ref{eq23})
and we also need $F_a^{'}$ to calculate $\phi[y_e,y_e]$.

From Eq.(\ref{eq21}) it follows that
\begin{eqnarray}
F_a(t)=\frac{\langle
a|\hat{x}(0)\hat{x}(t)|a\rangle}{\hbar},\label{eq24}
\end{eqnarray}
where $\hat{x}(t)$ is the Heisenberg representation of $\hat{x}$.
The real and imaginary parts of $F_a$ are
\begin{eqnarray}
F_a^{'}(t)=\frac{\langle a| \{\hat{x}(0),\hat{x}(t)\}| a
\rangle}{2\hbar},
\end{eqnarray}
and
\begin{eqnarray}
F_a^{''}(t)=\frac{\langle
a|[\hat{x}(0),\hat{x}(t)]|a\rangle}{2i\hbar},
\end{eqnarray}
where $\{.\}$ is the anticomutator and $[.]$ is the comutator. Thus,
$F_a^{'}$ and $F_a^{''}$ are, respectively, the quantum analogs of
the classical correlation and response functions of Section II.

%%%%%%%%%%%%%%%%%%%%%%%%%%%%%%%%%%%%%%%%%%%%%%%%%%%%%%%%%%%%%%
\subsection{Semiclassical Expressions for Correlation Functions.}

In this section we obtain semiclassical formulas for $F_a^{'}(t)$
and $F_a^{''}(t)$. We write
\begin{eqnarray}
F_a^{'}(t) = \frac{1}{2\hbar} \sum_b\langle
b|\{\hat{x}(0),\hat{x}(t)\}\hat{\rho}|b\rangle =
\frac{1}{2\hbar}Tr(\hat{f}(t)\hat{\rho}),\label{eq25}
\end{eqnarray}
where $\hat{\rho}=|a\rangle\langle a|$ is the microcanonical
distribution of the chaotic system and $\hat{f} =
\{\hat{x}(0),\hat{x}(t)\}$. To calculate $F''_a$ we take
$\hat{f}=-i[\hat{x}(0),\hat{x}(t)]$. Using the Wigner-Weyl
representation \cite{wigner} the trace can be written as
\begin{eqnarray}
Tr[\hat{f}(t)\hat{\rho}]=\int\frac{\mathrm{d}^2 q\mathrm{d}^2 p}
{(2\pi\hbar)^2}f(q,p;t)W(q,p) \label{eq26}
\end{eqnarray}
where
\begin{eqnarray}
f(q,p;t)=\int_{-\infty}^{\infty}\mathrm{d}u \, e^{
\frac{i}{\hbar}p\cdot u}\langle
q-u/2|\hat{f}(t)|q+u/2\rangle.\label{eq27}
\end{eqnarray}
is the Weyl transformation (or symbol) of $\hat{f}(t)$ and $W(q,p)$
is the Wigner function of $\hat{\rho}$. For
$\hat{f}(t)=-i[\hat{x}(0),\hat{x}(t)]$ we have
\begin{eqnarray}
f(q,p;t)&=&-i\int_{-\infty}^{\infty}\mathrm{d}u \, e^{ ip\cdot
u/\hbar} \, \langle
q-u/2|(\hat{x}\hat{x}(t)-\hat{x}(t)\hat{x})|q+u/2\rangle
\nonumber \\
&=&-i\int_{-\infty}^{\infty}\mathrm{d}u \, e^{ ip\cdot u/\hbar} \,
(-u_x)\langle q-u/2|\hat{x}(t)|q+u/2\rangle
\nonumber \\
&=&\hbar\frac{\partial}{\partial
p_x}\left\{\int_{-\infty}^{\infty}\mathrm{d} u \, e^{ ip\cdot
u/\hbar} \,\langle
q-u/2|\hat{x}(t)|q+u/2\rangle\right\} \nonumber \\
 &=& \hbar\frac{\partial}{\partial p_x} \left\{
\int_{-\infty}^{\infty}\mathrm{d} u\, e^{ ip\cdot u/\hbar} \,
\int_{-\infty}^{\infty}\mathrm{d} v\, v_x
K^*(v,q-u/2;t)K(v,q+u/2;t)\right\} \label{eq28}
\end{eqnarray}
where $u$ and $v$ represent coordinates of the chaotic system and
$K(v,v';t)=\langle v|e^{-i \hat{H}_c t/\hbar}|v'\rangle$ is the
propagator.

We now replace the propagators by their semiclassical expressions
$\tilde{K}$ and do the integrals by the stationary phase
approximation. The stationary phase condition shows that the most
important contributions come from the trajectories starting at
$q-u/2$ (for $K^*$) and $q+u/2$ (for $K$) and arriving at $v$ in the
time $t$ such that
\begin{eqnarray}
\nabla_{v}R_k(v,q-u/2)-\nabla_{v}R_l(v, q+u/2)=0,\label{eq30}
\end{eqnarray}
where $R_k$ and $R_l$ are Hamilton's principal functions coming from
the phases in $\tilde{K}$ and $\tilde{K}^*$. Since $\nabla_{v}
R_i(v,v')$ gives the final momentum, (\ref{eq30}) imposes that the
final momenta of the two trajectories must be equal. Since the final
positions are also equal, the two trajectories must be identical.
Thus,
\begin{eqnarray}
\int_{-\infty}^{\infty}\mathrm{d}v\, v_x \tilde{K}^*(
v,q-u/2;t)\tilde{K}( v,q+ u/2;t)\approx x(q,p;t)\delta(u),
\label{eq31}
\end{eqnarray}
where $x(q,p;t)$ is the coordinate $x$ of the stationary trajectory.

Using (\ref{eq31}) in (\ref{eq28}) we find
\begin{eqnarray}
f(q,p;t)=\hbar\frac{\partial}{\partial p_x} x (q,p;t) =
\hbar\{x(0),x(t)\}.\label{eq32}
\end{eqnarray}
since
\begin{eqnarray}
\{x(0),x(t)\}=\frac{\partial x(0)}{\partial x(0)}\frac{\partial
  x(t)}{\partial p_x(0)}-\frac{\partial x(0)}{\partial
  p_x(0)}\frac{\partial x(t)}{\partial x(0)}=\frac{\partial
  x(t)}{\partial p_x(0)}.\label{eq33}
\end{eqnarray}

For $\hat{f}(t)=\{\hat{x}(0),\hat{x}(t)\}$, we find
\begin{eqnarray}
f(q,p;t)&=&
\int_{-\infty}^{\infty}\mathrm{d}u\exp\left(\frac{i}{\hbar} p\cdot
u\right)2q_x\langle q-u/2|\hat{X}(t)|q+u/2\rangle \nonumber
 \\ &\approx& 2q_x\, x(q,p;t)=2x(0)x(t). \label{eq35}
\end{eqnarray}

The semiclassical limit of the Wigner function
\begin{eqnarray}
W(q,p;E)=\frac{1}{\hbar}\int_{-\infty}^{\infty}\mathrm{d}t
e^{iEt/\hbar}\int\mathrm{d}u\exp\left(\frac{i}{\hbar}p\cdot u
\right)K(q+u/2,q-u/2;t).\label{eq37}
\end{eqnarray}
was obtained by Berry \cite{berry2} and can be written as
\begin{eqnarray}
W(q,p;E)\approx\delta(E-H(q,p))+W_1(q,p;E). \label{eq38}
\end{eqnarray}
The first term is the classical micro-canonical distribution and the
second, $W_1$, is given by classical periodic orbits corrections to
the classical function. These periodic orbits have energy $E=E_a$,
corresponding to the eigenstate $|a\rangle$ of the microcanonical
quantum distribution. Using (\ref{eq38}) we write
\begin{eqnarray}
Tr[\hat{f}(t)\hat{\rho}] &\approx
\displaystyle{\frac{1}{(2\pi\hbar)^2}} & \left[ \int\mathrm{d}q
\mathrm{d} p f(q,p;t)\delta(E-H(q,p)) \, + \int\mathrm{d}q
\mathrm{d} p
f(q,p;t)W_1(q,p;E) \right]\nonumber \\
& & \equiv \frac{1}{(2\pi\hbar)^2} \left[f^0 + f^1\right].
\label{eq39}
\end{eqnarray}
When $f(q,p;t)$ is replaced by the semiclassical expressions for the
anticomutator and comutator, the first term of (\ref{eq39}) becomes,
except for a normalization, the classical expressions for the
response and correlation functions respectively. Following
\cite{berry2} the second term of (\ref{eq39}) becomes
\begin{eqnarray}
f^1(E,t) \approx \sum_jA_j\cos{(S_j(E)/\hbar+\gamma_j)}\oint
\mathrm{d}\tau f(q_j(\tau),p_j(\tau);t).\label{eq40}
\end{eqnarray}
where, $A_j$ depends on the stability of $j$-periodic orbit,
$S_j(E)$ is its action, $\gamma_j$ is the Maslov index and the
integral is calculated over a period of the $j$-orbit. Analogous
results can be found, for example, in \cite{eckart}.

Furthermore \cite{berry2},
\begin{equation}
\begin{array}{ll}
Tr(\hat{\rho})&=\displaystyle{\frac{1}{(2\pi\hbar)^2}} \left[
\int\mathrm{d}q\mathrm{d}p
\,\delta(E-H(q,p))+n_q(E;\hbar) \right] \\ \\
&=\displaystyle{\frac{1}{(2\pi\hbar)^2}}[n_c(E)+ n_q(E;\hbar)]\\
\label{eq42}
\end{array}
\end{equation}
where the first term is the classical density of states and the
second term is known as Gutzwiller's trace formula. Thus,
\begin{equation}
\begin{array}{ll}
\displaystyle{\frac{Tr[\hat{f}(t)\hat{\rho}]}{Tr[\hat{\rho}]}} &
\approx \displaystyle{\frac{f^0(E;t)+f^1(E;t)}{n_c(E)+n_q(E;\hbar)}}
\\ \\ & \approx \langle f(q,p;t)\rangle_{cl.}\left[1-
\frac{n_q(E;\hbar)}{n_c(E)}\right]+\frac{f^1(E;t)}{n_c(E)}\left[1-
\frac{n_q(E;\hbar)}{n_c(E)}\right] \label{eq47}
\end{array}
\end{equation}
where $\langle f(q,p;t)\rangle_{cl.}=f^0(E,t)/n_c(E)$.
%
%\begin{eqnarray}
%\langle f(q,p;t)\rangle_{cl.}=\int\mathrm{d}q
%\mathrm{d}p\,f(q,p;t)\delta(E-H(q,p))\Big/
%\int\mathrm{d}q\mathrm{d}p\,\delta(E-H(q,p)). \label{eq48}
%\end{eqnarray}
%

Finally, we can calculate the semiclassical expressions for
$F_a^{''}(t)$ and $F_a^{'}(t)$. From (\ref{eq32}), (\ref{eq35}) and
(\ref{eq47})
\begin{eqnarray}
F_a^{'}(t)&\approx& \frac{\langle
  x(0)x(t)\rangle_{cl.}}{\hbar}
\left[1-\frac{n_q(E;\hbar)}{n_c(E)}\right]
\nonumber \\
&+&\frac{1}{\hbar n_c(E)}\sum_j A_j
\cos\left(S_j(E)/\hbar+\gamma_j\right) \oint\mathrm{d}\tau
x_j(\tau)x_j(\tau+t).\label{eq49}
\end{eqnarray}
and
\begin{eqnarray}
F_a^{''}(t)=\frac{\langle\hat{O}(t)\rangle}{2\hbar}&\approx&\frac{\langle
  \{x(0),x(t)\}\rangle_{cl.}}{2}
  \left[1-\frac{n_q(E;\hbar)}{n_c(E)}\right]
\nonumber \\
&+&\frac{1}{2 n_c(E)}\sum_j A_j \cos\left(S_j(E)/\hbar+\gamma_j\right)
\oint\mathrm{d}\tau \{x_j(\tau),x_j(\tau+t)\}.\nonumber \\
\,\label{eq50}
\end{eqnarray}

Both these semiclassical expressions are given by their classical
counterparts multiplied by a correction to their amplitudes, given
by $n_q(E;\hbar)/n_c(E)$, plus a correction from periodic orbits.
The temporal dependence of the first is given solely by the
classical dynamics and it decays exponentially. The second  term,
however, is a sum of oscillating functions and carries the temporal
dependence characteristic of the chaotic system. As a final remark
we note in \cite{esposito1} these functions were calculated using
random matrix theory.

%%%%%%%%%%%%%%%%%%%%%%%%%%%%%%%%%%%%%%%%%%%%%%%%%%%%%%%%%%%%%%
\subsection{The Superpropagator}

The semiclassical expressions for $F_a^{'}(t)$ and $F_a^{''}(t)$
allows us to solve the equations of motion (\ref{eq22}) and
(\ref{eq23}) and to calculate the superpropagator,
Eq.(\ref{superprop}). In order to have explicit formula, we shall
consider only the zero order approximation $F_a^{''}(t) \approx
\phi_{xx}(t)/2$. Substituting it in (\ref{eq22}) and (\ref{eq23})
and integrating by parts we get
\begin{eqnarray}
\ddot{r}_e(t)+\Omega^2_0\, r_e(t)+\frac{\gamma^2}{m}\int_0^t\mathrm{d}s
F(t-s)\dot{r}_e(s)&=&-\frac{\gamma^2}{m}r_e(0)F(t), \label{eq-3}\\
\ddot{y}_e(t)+\chi^2_0\, y_e(t)-\frac{\gamma^2}{m}\int_0^t\mathrm{d}s
F(t-s)\dot{y}_e(s)&=&\frac{\gamma^2}{m}y_e(0)F(t), \label{eq-4}
\end{eqnarray}
where $\Omega_0^2=\omega_0^2-\gamma^2F(0)/m$, $\chi^2_0=\omega_0^2+
\gamma^2F(0)/m$ and
\begin{eqnarray}
F(t)=\int\mathrm{d}s\phi_{xx}(t-s),\label{eq-5}
\end{eqnarray}
as in the classical case (eq.(\ref{eq17})).

In Appendix B we solve these equations by the method of Laplace
transforms. We find that the solutions involve two very different
time scales. The shortest time scale is relevant only for times much
smaller than $1/\omega_0$, the period of the decoupled oscillator.
For times of the order of $1/\omega_0$ these terms can be discarded
as transients. Here we shall adopt this approximation and keep only
the terms that are significant for times of $1/\omega_0$. In this
case we show that (\ref{eq-3}) and (\ref{eq-4}) can be written as
\begin{eqnarray}
\ddot{r}_e(t)+2\Lambda\dot{r}_e(t)+\Omega_0^2r_e(t)=0, \quad
\Omega_0^2\gg\Lambda^2
,\label{eq-6} \\
\ddot{y}_e(t)-2\Lambda\dot{y}_e(t)+\chi_0^2y_e(t)=0,\quad
\chi_0^2\gg\Lambda^2, \label{eq-7}
\end{eqnarray}
where
\begin{eqnarray}
\Lambda=\frac{\gamma^2}{2m}\lim_{t\to\infty}\int_0^{t}\mathrm{d}s
F(t-s).\label{eq-8}
\end{eqnarray}
Approximating $\Omega_0 \approx \chi_0 \approx \omega_0$ we find
\begin{eqnarray}
r_e(t)=e^{-\Lambda
  t}\left\{\frac{\sin{[\omega_0(T-t)]}}{\sin{(\omega_0 T)}}r(0)+
e^{\Lambda T} \frac{\sin{(\omega_0 t)}}{\sin{(\omega_0
  T)}}r(T)\right\},\nonumber \\
y_e(t)=e^{\Lambda
  t}\left\{\frac{\sin{[\omega_0(T-t)]}}{\sin{(\omega_0 T)}}y(0)+
e^{-\Lambda T} \frac{\sin{(\omega_0 t)}}{\sin{(\omega_0
  T)}}y(T)\right\}.\label{eq-10}
\end{eqnarray}

Within this approximation we can calculate the real part $\tilde{S}$
of the effective action, Eq.(\ref{sefreal}). We obtain
\begin{eqnarray}
\tilde{S}[r_e,y_e]&=&[m\omega_0^2
K(T)-m\Lambda]r(T)y(T)+[m\omega_0^2 K(T)
+m\Lambda]r(0)y(0) \nonumber \\
& &-m\omega_0^2L(T)r(0)y(T)-m\omega_0^2N(T)
r(T)y(0),\label{eq-13}
\end{eqnarray}
where,
\begin{eqnarray}
K(T)&=&\frac{1}{\omega_0}\frac{\cos(\omega_0 T)}
{\sin(\omega_0 T)}, \label{eq-14}\\
L(T)&=&\frac{1}{\omega_0}\frac{e^{-\Lambda T}}{\sin(\omega_0 T)},
\label{eq-15}\\
N(T)&=&\frac{1}{\omega_0}\frac{e^{\Lambda T}}{\sin(\omega_0 T)}.\label{eq-16}
\end{eqnarray}
For the imaginary part $\phi$ of the effective action,
Eq.(\ref{sefimag}), we get
\begin{eqnarray}
\phi[y_e,y_e]=\gamma^2\int_0^T\mathrm{d}t\,y_e(t)\int_0^t\mathrm{d}s\,
F_a^{'}(t-s)y_e(s).\label{eq-17}
\end{eqnarray}
Using again only the zero order term of the semiclassical expression
(\ref{eq49}) for $F_a^{'}(t)$ and the fact that the classical
correlation function decays exponentially, as in (\ref{eq-8}), we
can approximate, for $t \sim 1/\omega_0$
\begin{eqnarray}
\int_0^t\mathrm{d}s\,F_a^{'}(t-s)y_e(s)\approx y_e(t)
\lim_{t\to\infty}\int_0^{t}\mathrm{d}s\,F_a^{'}(t-s) \equiv y_e(t)
\frac{B'}{\hbar}, \label{eq-20}
\end{eqnarray}
which implies that
\begin{equation}
\begin{array}{ll}
\phi[y_e,y_e] &= \displaystyle{\frac{\gamma^2B'}{\hbar}}
\int_0^T\mathrm{d}t\,y^2_e(t) \\
&=\displaystyle{\frac{\gamma^2B'}{\hbar}}[A(T)y^2(T)+B(T)y(T)y(0) +
C(T)y^2(0)],\label{eq-22}
\end{array}
\end{equation}
where
\begin{eqnarray}
A(T)&=&\frac{[-\omega_0^2e^{-2\Lambda
      T}+(\Lambda^2+\omega_0^2)-\Lambda^2\cos{(2\omega_0
      T)-\Lambda\omega_0\sin{(2\omega_0
      T)}}]}{4\Lambda(\omega_0^2+\Lambda^2)\sin^2{(\omega_0 T)}},
      \label{eq-x1} \\
B(T)&=&\frac{[-\omega_0^2\sinh{(\Lambda
      T)}\cos{(\omega_0 T)}+\Lambda\omega_0\cosh{(\Lambda T)}\sin{(\omega_0
      T)}]}{\Lambda(\omega_0^2+\Lambda^2)\sin^2{(\omega_0 T)}},
      \label{eq-x2} \\
C(T)&=&\frac{[\omega_0^2e^{2\Lambda
      T}-(\Lambda^2+\omega_0^2)+\Lambda^2\cos{(2\omega_0
      T)-\Lambda\omega_0\sin{(2\omega_0
      T)}}]}{4\Lambda(\omega_0^2+\Lambda^2)\sin^2{(\omega_0 T)}}.
      \label{eq-x3}
\end{eqnarray}

Finally, putting everything together we get
\begin{eqnarray}
\lefteqn{J(r(T),y(T),r(0),y(0))=G(T,0)\exp{\left\{\frac{i}{\hbar}\tilde{S}
[r_e,y_e]\right\}}\exp{\left\{-\frac{1}{\hbar}\phi[y_e,y_e]\right\}}
}\nonumber \\
&=&G(T,0)\exp{\left\{\frac{i}{\hbar}\tilde{K}_2(T)r(T)y(T)\right\}}
\exp{\left\{-\frac{1}{\hbar}\tilde{A}(T)y^2(T)\right\}}\times \nonumber \\
&
&\exp{\left\{\frac{i}{\hbar}\left[\tilde{K}_1(T)r(0)y(0)-\tilde{L}(T)r(0)y(T)
-\tilde{N}(T)r(T)y(0)\right]\right\}}\times \nonumber \\
& &\exp{\left\{-\frac{1}{\hbar}\tilde{B}(T)y(T)y(0)-\frac{1}{\hbar}
\tilde{C}(T)y^2(0)\right\}} \label{eq-24}
\end{eqnarray}
where
\begin{eqnarray}
\tilde{K}_{1,2}(T)=m\omega_0^2K(T)\pm m\Lambda
\qquad&\tilde{L}(T)=m\omega_0^2L(T)&\qquad
\tilde{N}(T)=m\omega_0^2N(T) \nonumber \\ \nonumber \\
\tilde{A}(T)=\frac{\gamma^2B'}{\hbar}A(T)\qquad&\tilde{B}(T)=
\frac{\gamma^2B'}{\hbar}B(T)&\qquad\tilde{C}(T)=\frac{\gamma^2B'}
{\hbar}C(T). \nonumber \\
\label{eq-23}
\end{eqnarray}

We finish this section with a comment about the physical situation
described by these calculations. Since we have considered only the
first terms of the semiclassical expressions for $F_a^{'}$ and
$F_a^{''}$ our results are valid only in the Ehrenfest time scale of
the chaotic system given by \cite{beenaker}
\begin{eqnarray}
t_E\sim\frac{1}{\lambda_L}\ln{\left(\frac{S_c}{\hbar}\right)},
\end{eqnarray}
where $\lambda_L$ is the Lyapunov exponent and $S_c$ is a typical
action of the chaotic system, for example the action of the shortest
periodic orbit. Approximating $\lambda_L$ by $\alpha$, we see that,
in order to observe effects for $t\sim 1/\omega_0$, we must have
\begin{eqnarray}
t_E\sim\frac{1}{\omega_0}=\frac{1}{\alpha}\ln{\left(\frac{S_c}{\hbar}\right)}
\Rightarrow
\frac{S_c}{\hbar}=\exp{\left(\frac{\alpha}{\omega_0}\right)}.
\label{eq-25}
\end{eqnarray}
For NS, $\alpha/\omega_0\sim 8$ and $S_c \approx 10$ which means
that $\hbar$ must be smaller than $10^{-3}$ for our results to be
valid.

%%%%%%%%%%%%%%%%%%%%%%%%%%%%%%%%%%%%%%%%%%%%%%%%%%%%%%%%%%%%%%
%%%%%%%%%%%%%%%%%%%%%%%%%%%%%%%%%%%%%%%%%%%%%%%%%%%%%%%%%%%%%%
\section{Applications}

The superpropagator allows us to study the time evolution of the
oscillator under the influence of the chaotic system. The reduced
density matrix satisfies
\begin{eqnarray}
\rho_o(r(T),y(T))=\int\mathrm{d}r(0)\mathrm{d}y(0)\,J(r(T),y(T),r(0),y(0))
\,\rho_o(r(0),y(0)). \label{eq-26}
\end{eqnarray}

In the following we calculate explicitly the propagation of two
different oscillator's states. These two applications are similar to
the ones presented by Caldeira and Leggett to study dissipation and
decoherence.

%%%%%%%%%%%%%%%%%%%%%%%%%%%%%%%%%%%%%%%%%%%%%%%%%%%%%%%%%%%%%%
\subsection{Propagation of a Gaussian State.}

For a Gaussian state
\begin{eqnarray}
\psi(z(0))=\frac{1}{(2\pi\sigma^2)^{1/4}}\,e^{\frac{i}{\hbar}pz(0)/\hbar}
e^{-z^2(0)/4\sigma^2} \label{eq-28}
\end{eqnarray}
the density matrix $\rho_o(z(0),z'(0))=\psi^*(z'(0))\psi(z(0))$ can
be written in terms of $r=(z+z')/2$ and $y=z-z'$ as
\begin{eqnarray}
\rho_o(r(0),y(0))=\frac{1}{(2\pi\sigma^2)^{1/2}}\,e^{\frac{i}{\hbar}py(0)}
e^{-r^2(0)/2\sigma^2}e^{-y^2(0)/8\sigma^2}. \label{eq-29}
\end{eqnarray}

Substituting (\ref{eq-24}) and (\ref{eq-29}) in (\ref{eq-26}) and
performing the integrals, we get
\begin{eqnarray}
\lefteqn{\rho_o(r(T),y(T))=G(T,0)\left[\frac{2\pi\hbar^2}{2\hbar
    \tilde{C}_1(T)+\sigma^2\tilde{K}_1^2(T)}\right]^{1/2}}\nonumber \\
&\times \exp{\left\{-
\frac{\tilde{N}^2(T)}{2[2\hbar\tilde{C}_1(T)+\sigma^2\tilde{K}_1^2(T)]}
\left(r(T)-\frac{p}{\tilde{N}(T)}\right)^2\right\}}&
\nonumber \\
&\times\exp{\left\{
-\left[\frac{\tilde{A}(T)}{\hbar}+\frac{\sigma^2
\tilde{L}^2(T)}{2\hbar^2}-\frac{(\sigma^2\tilde{K}_1(T)\tilde{L}(T)-\hbar
\tilde{B}(T))^2}{2\hbar^2[2\hbar\tilde{C}_1(T)+\sigma^2\tilde{K}_1^2(T)]}
\right]y^2(T)\right\}}& \nonumber \\
&\times\exp\left\{\frac{i}{\hbar}\tilde{K}_2(T)r(T)y(T)
-\frac{i}{\hbar}\frac{(\sigma^2\tilde{K}_1(T)\tilde{L}(T)
-\hbar\tilde{B}(T))}{(2\hbar\tilde{C}_1(T)+\sigma^2\tilde{K}_1^2(T))}
\tilde{N}(T)
\left(r(T)-\frac{p}{\tilde{N}(T)}\right)y(T)\right\}&, \nonumber \\
\label{eq-30}
\end{eqnarray}
where,
\begin{eqnarray}
\tilde{C}_1(T)=\tilde{C}(T)+\frac{\hbar}{8\sigma^2}. \label{eq-31}
\end{eqnarray}

For $y=z-z'=0$, $\rho_o$ becomes the probability density. After
normalizing we obtain
\begin{eqnarray}
\rho_o(r(T),0)&=&\left[\frac{\tilde{N}^2(T)}{2\pi[2\hbar\tilde{C}_1(T)+
\sigma^2\tilde{K}_1^2(T)]}\right]^{1/2}\nonumber \\
&\times&\exp{\left\{-\frac{\tilde{N}^2(T)}
{2[2\hbar\tilde{C}_1(T)+\sigma^2\tilde{K}_1^2(T)]}\left(r(T)-\frac{p}
{\tilde{N}(T)}\right)^2\right\}}. \label{eq-33}
\end{eqnarray}
Eq.(\ref{eq-33}) represents a Gaussian packet whose center follows
the trajectory
\begin{eqnarray}
r(T)=\frac{p}{\tilde{N}(T)}=\frac{p}{m\omega_0}e^{-\Lambda
  T}\sin{(\omega_0 T)}. \label{eq-34}
\end{eqnarray}
The dissipative effect due to the interaction with the chaotic
system is explicit. The same behavior was obtained by Caldeira and
Leggett \cite{caldeira1} using a thermal bath with many degrees of
freedom. Eq.(\ref{eq-34}) represents the trajectory of a weakly
damped harmonic oscillator. The critical and strongly damped cases
cannot be described by this formalism because of the weak coupling
regime adopted.

The width of the evolved packet is given by
\begin{eqnarray}
\sigma^2(T)&=&\frac{\sigma^2\tilde{K}_1^2(T)+2\hbar\tilde{C}_1(T)}
{\tilde{N}^2(T)} .
\end{eqnarray}
After some algebra we can show that
\begin{eqnarray}
\sigma^2(T)&=& \sigma^2\bigg\{\frac{(1+\epsilon^2)e^{-2\Lambda
T}}{1+ \epsilon^2} \nonumber \\
&+&\frac{\Gamma[1-e^{-2\Lambda T}(1+2\epsilon\sin{(\omega_0
T)}\cos{(\omega_0 T)}+2\epsilon^2 \sin^2{(\omega_0
T)})]}{1+\epsilon^2}\bigg\}, \label{eq-35}
\end{eqnarray}
where
\begin{eqnarray}
\epsilon=\frac{\Lambda}{\omega_0}\qquad\mathrm{and}\qquad\Gamma=\frac{E_c(0)}
{\hbar\omega_0}. \label{eq-36}
\end{eqnarray}
The expression above for $\Gamma$ comes from the following
considerations: from $\tilde{C}_1(T)$ it follows that
\begin{eqnarray}
\Gamma=\frac{\gamma^2B'}{\hbar m\omega_0\Lambda}.
\end{eqnarray}
Using Eqs.(\ref{eq-8}) and (\ref{eq17}) and the relation
\cite{weiss}
\begin{eqnarray}
\langle p_x(0)x(t)\rangle_e=-\frac{\partial}{\partial t}\langle
x(0)x(t)\rangle_e
\end{eqnarray}
we obtain
\begin{eqnarray}
\Lambda=\frac{\gamma^2 B'}{m E_c(0)},
\end{eqnarray}
which leads directly to (\ref{eq-36}). Notice that, due to
(\ref{eq-25}), $\Gamma\gg 1$ is the only possibility.

Fig.3 shows that $\sigma^2(T)$ for $\Gamma=1$, $0.5$ and $2.0$.
These curves can be well fitted by the simpler expression
\begin{eqnarray}
\sigma^2(T)=\sigma^2[e^{-2\Lambda T}+\Gamma(1-e^{-2\Lambda T})],
\label{eq-37}
\end{eqnarray}
which, for $t\sim 1/\omega_0$, can be written as
\begin{eqnarray}
\sigma^2(T)=\sigma^2[1+(\Gamma-1)2\Lambda T]. \label{eq-38}
\end{eqnarray}
We see that $\sigma^2(\Gamma-1)\Lambda$ plays the role of a
diffusion constant. Fig.3 also shows that $\Gamma$ controls the
increase or decrease of $\sigma^2(T)$. In the present case,
$\sigma^2(T)$ can only increase because of the constraint $\Gamma\gg
1$. In the Caldeira-Leggett model, on the other hand, the width can
also decrease if the the temperature is very low.

%%%%%%%%%%%%%%%%%%%%%%%%%%%%%%%%%%%%%%%%%%%%%%%%%%%%%%%%%%%%%%
\subsection{Superposition of Two Gaussian States.}

We now consider an initial state consisting of two Gaussian
wave-packets, one at the origin and one centered at $z(0)=q_0$:
\begin{eqnarray}
\psi(z(0))&=&N^{1/2}[\psi_1(z(0))+\psi_2(z(0))]\nonumber \\
&=&N^{1/2}\left\{\exp{\left[
-\frac{z^2(0)}{4\sigma^2}\right]}+\exp{\left[-\frac{(z(0)-q_0)^2}
{4\sigma^2}\right]}\right\}. \label{eq-39}
\end{eqnarray}
The density matrix is given by
\begin{eqnarray}
\rho_o(z(0),z'(0))=&N&[\rho_{11}(z(0),z'(0))+\rho_{22}(z(0),z'(0))\nonumber \\
&+&\rho_{12}(z(0),z'(0))+\rho_{21}(z(0),z'(0))]. \label{eq-40}
\end{eqnarray}
with $\hat{\rho}_{ij}=|\psi_i\rangle\langle\psi_j|$. The time
evolution of $\rho_o$ can again be calculated with Eq.(\ref{eq-26}).
The result, for $y=z-z'=0$ is
\begin{eqnarray}
\rho_{11}(r(T),0)&=&\frac{1}{2[1+h(T)]}\left(\frac{\tilde{N}^2(T)}
{\pi\tilde{f}(T)}\right)^{1/2}
\exp{\left\{-\frac{\tilde{N}^2(T)}{\tilde{f}(T)}r^2(T)\right\}},
\label{eq-41} \\
\rho_{22}(r(T),0)&=&\frac{1}{2[1+h(T)]}\left(\frac{\tilde{N}^2(T)}
{\pi\tilde{f}(T)}\right)^{1/2}
\exp{\left\{-\frac{\tilde{N}^2(T)}{\tilde{f}(T)}\left[r(T)-Q(T)\right]^2
\right\}}, \label{eq-42} \\
\rho_{12}(r(T),0)&+&\rho_{21}(r(T),0) \, = \nonumber \\
& &\frac{1}{2[1+h(T)]}\left(
\frac{\tilde{N}^2(T)}{\pi\tilde{f}(T)}\right)^{1/2}\exp{
\left[-\frac{q^2_0}{8\sigma^2}g(T)\right]}\exp{\left\{-\frac
{\tilde{N}^2(T)}{\tilde{f}(T)}r^2(T)\right\}} \nonumber \\
&\times&\exp{\left\{-\frac{\tilde{N}^2(T)}{\tilde{f}(T)}\left[r(T)-Q(T)
\right]^2\right\}}\nonumber \\
&\times&2\cos\left\{\frac{\hbar\tilde{N}^2(T)}
{4\sigma^2\tilde{f}(T)\tilde{K}_1(T)}
\left[\left(r(T)-Q(T)\right)^2-r^2(T)\right]\right\}, \label{eq-43}
\end{eqnarray}
where
\begin{eqnarray}
\tilde{f}(T)&=&2[2\hbar\tilde{C}_1(T)+\sigma^2\tilde{K}_1^2(T)],
\label{eq-44} \\
h(T)&=&\exp{\left\{-\frac{q^2_0}{8\sigma^2}\left[1+g(T)\right]\right\}},
  \label{eq-45} \\
Q(T)&=&\frac{\tilde{K}_1(T)}{\tilde{N}(T)}q_0, \label{eq-46} \\
g(T)&=&\frac{2\hbar\tilde{C}(T)}{2\hbar\tilde{C}_1(T)+\sigma^2
\tilde{K}_1^2(T)}.
\label{eq-47}
\end{eqnarray}

The interference term can also be rewritten as
\begin{eqnarray}
\rho_{12}(r(T),0)&+&\rho_{21}(r(T),0)=2\cos{\left[a(T)((r(T)-Q(T))^2-r^2(T))
\right]}\nonumber \\
&\times&\rho_1^{1/2}(r(T),0)\rho_2^{1/2}(r(T),0)\exp{\left[-\frac{q^2_0}
{8\sigma^2}g(T)\right]}. \label{eq-48}
\end{eqnarray}
Eq.(\ref{eq-48}) shows that the interference is attenuated by
$\exp{[-(q^2_0/8\sigma^2)g(T)]}$. Eq.(\ref{eq-48}) is very similar
to the expression obtained by Caldeira and Leggett \cite{caldeira2},
although there is no temperature dependence in $g(T)$, which can be
written as
\begin{eqnarray}
g(T)=\frac{\Gamma\,b(T)}{(1+\epsilon^2)+\Gamma\,b(T)}, \label{eq-49}
\end{eqnarray}
with
\begin{eqnarray}
b(T)=e^{2\Lambda T}-1-2\epsilon\sin{(\omega_0 T)}\cos{(\omega_0
  T)}-2\epsilon^2\sin^2{(\omega_0 T)} \label{eq-50}
\end{eqnarray}
and $\epsilon=\Lambda/\omega_0$. We note that the asymptotic limits
\begin{eqnarray}
g(T=0)=0\qquad\mathrm{and}\qquad g(T\rightarrow\infty)\rightarrow 1
\label{eq-51}
\end{eqnarray}
are the same as those in the Caldeira-Leggett model.

Fig.4 shows $g(T)$ for $\Gamma=10$. In the regime $\Gamma\gg 1$, we can
approximate $g(T)$ by
\begin{eqnarray}
g(T)=\frac{2\Gamma\Lambda T}{1+2\Gamma\Lambda T}. \label{eq-52}
\end{eqnarray}
This simplified expression helps to estimate of the decoherence
time. For example, with (\ref{eq-52}), we can estimate the time $T'$
such that
\begin{eqnarray}
\exp{\left[-\frac{q^2_0}{8\sigma^2}g(T')\right]}\sim
10^{-3}. \label{eq-53}
\end{eqnarray}
Defining $n\equiv q^2_0/8\sigma^2$ (the number of quanta
$\hbar\omega_0$ of the wave packet centered at $q_0$), we get
\begin{eqnarray}
\left[\frac{n-\ln(10)}{3\ln(10)}\right]2\Gamma\Lambda
T'=2\tilde{n}\Gamma\Lambda T'=1\Rightarrow
T'=\frac{1}{2\tilde{n}\Gamma\Lambda}. \label{eq-54}
\end{eqnarray}
Since we are interested in the situation where $n \gg 1$ and $\Gamma
\gg 1$, we find that the decoherence time is much smaller than the
time scale where dissipation takes place, i.e.,  $T' \ll 1/\Lambda$.

%%%%%%%%%%%%%%%%%%%%%%%%%%%%%%%%%%%%%%%%%%%%%%%%%%%%%%%%%%%%%%
%%%%%%%%%%%%%%%%%%%%%%%%%%%%%%%%%%%%%%%%%%%%%%%%%%%%%%%%%%%%%%
\section{Discussion and Conclusions}

We have made two important assumptions in our calculation of the
superpropagator. The first of these assumptions, the weak coupling
regime, was important to reduce the path integral to a quadratic
form in the oscillator variables. The second assumption was the
semiclassical regime of the chaotic system. This was essential to
establish the connection between the coupling in the influence
functional and the classical correlation and response functions that
enter in the classical description of the system. The use of these
classical functions make the importance of the chaotic dynamics
explicit and show that the time scales obtained classically are
important ingredients to describe dissipation. In particular, the
exponential decay of correlations happens in a time scale much
shorter than the natural period of the oscillator. The time of
correlation loss plays the role of the microscopic time scale in the
Brownian motion, which is much shorter than the macroscopic one
\cite{reif}. Moreover, the exponential decay of the classical
correlations is what makes dissipation possible in the present
treatment. The corrections due to periodic orbits have not been
explored here and the importance of their contribution to
dissipation and decoherence is not clear at this point.

The effective dynamics we obtained, expressed in (\ref{eq-24}), is
analogous to the Caldeira-Leggett theory in the limit of high
temperatures and weak damping \cite{caldeira1,caldeira2}. For
example, the diffusion constant in (\ref{eq-38})  can be written,
for $\Gamma\gg1$, as
\begin{eqnarray}
\sigma^2\Gamma\Lambda=\frac{E_c(0)}{2m\omega_0^2}\Lambda,
\end{eqnarray}
which should be compared with
\begin{eqnarray}
D=\frac{k_B T}{m\omega_0^2}\Lambda
\end{eqnarray}
for the Brownian motion. Therefore, $E_c(0)$ plays the role of $k_B
T$. From Fig.4, $\Gamma$ seems to play the role of $k_B
T/\hbar\omega_k$ since it controls the behavior of $\sigma^2(T)$.
However, despite this close analogy between the two models, our
results are valid only for short times since they are limited by
Ehrenfest time and perturbation theory.

In summary, we have shown, using Feynman-Vernon approach, that a
chaotic system with two degrees of freedom can induce dissipation
and decoherence in a simple quantum system when weakly coupled to
it. The formalism we have chosen allows us a close analogy with the
many body formulation of the Caldeira-Leggett model. The most
important quantities in the formalism, the correlation and response
functions, are obtained directly from the dynamics, and not from
phenomenological assumptions as in the Caldeira-Legget model. In our
approach we have used simple classical approximations and discarded
all periodic orbits corrections. The effects of these corrections
are certainly worth studying.

%%%%%%%%%%%%%%%%%%%%%%%%%%%%%%%%%%%%%%%%%%%%%%%%%%%%%%%%%%%%%%
%%%%%%%%%%%%%%%%%%%%%%%%%%%%%%%%%%%%%%%%%%%%%%%%%%%%%%%%%%%%%%
\begin{appendix}

\section{The Stationary Phase Approximation}

In this appendix we solve the path integral Eq.(\ref{pathint}) by
the stationary phase approximation. Let $(r_e(t),y_e(t))$ be the
stationary path and
\begin{eqnarray}
&r(t)=r_e(t)+\delta r(t)=r_e(t)+\epsilon_1\tilde{r}(t)& \\
&y(t)=y_e(t)+\delta y(t)=y_e(t)+\epsilon_2\tilde{y}(t)&,
\end{eqnarray}
be a neighboring path with $\tilde{r}(T)=\tilde{r}(0)=0$ and
$\tilde{y}(T)=\tilde{y}(0)=0$.

The stationary path is obtained from the condition
\begin{equation}
\begin{array}{ll}
\Delta \tilde{S} & \equiv
\tilde{S}[r_e(t)+\epsilon_1\tilde{r}(t),y_e(t)+\epsilon_2
\tilde{y}(t)]-\tilde{S}[r_e(t),y_e(t)] \\
 &= \epsilon_1\frac{d\Delta\tilde{S}}{d\epsilon_1}+\epsilon_2\frac{d\Delta
\tilde{S}}{d\epsilon_2} = 0
\end{array}
\end{equation}
We find
\begin{eqnarray}
\frac{d\Delta\tilde{S}}{d\epsilon_1}
=-\int_0^T\mathrm{d}t\tilde{r}(t)\left\{m[\ddot{y}_e(t)+\omega_0^2
y_e(t)]-2\gamma^2\int_0^t\mathrm{d}sF_a{''}(s-t)y_e(s)\right\},
\end{eqnarray}
and
\begin{eqnarray}
\frac{d\Delta\tilde{S}}{d\epsilon_2}
=-\int_0^T\mathrm{d}t\,\tilde{y}(t)\left\{m[\ddot{r}_e(t)+\omega_0^2
r_e(t)]-2\gamma^2\int_0^t\mathrm{d}sF_a{''}(t-s)r_e(s)\right\},
\end{eqnarray}
where we used $\tilde{r}(T)=\tilde{r}(0)=0$,
$\tilde{y}(T)=\tilde{y}(0)=0$ and
$\int_0^T\mathrm{d}t\int_0^T\mathrm{d}s=2\int_0^T\mathrm{d}t\int_0^t
\mathrm{d}s$.

Therefore, the equations of motion for the stationary path are given
by
\begin{equation}
\ddot{y}_e(t)+\omega_0^2y_e(t)-\frac{2\gamma^2}{m}
\int_0^t\mathrm{d}sF_a{''}(s-t) y_e(s)=0\label{yclas}
\end{equation}
and
\begin{equation}
\ddot{r}_e(t)+\omega_0^2r_e(t)-\frac{2\gamma^2}{m}
\int_0^t\mathrm{d}sF_a{''}(t-s) r_e(s)=0.\label{rclas}
\end{equation}

Expanding $\phi$, Eq.\ref{sefimag}, around the stationary path we
find
\begin{eqnarray}
\lefteqn{\phi[r_e(t)+\epsilon_1\tilde{r}(t),y_e(t)+\epsilon_2
\tilde{y}(t)]=} \nonumber \\
& &=\frac{1}{2}\gamma^2\int_0^T\mathrm{d}t\int_0^T\mathrm{d}s[y_e(t)+
\epsilon_2
\tilde{y}(t)][y_e(s)+\epsilon_2\tilde{y}(s)]F_a{'}(t-s) \nonumber \\
& &=\frac{1}{2}\gamma^2\int_0^T\mathrm{d}t\int_0^T\mathrm{d}sy_e(t)y_e(s)
F_a{'}(t-s) + \nonumber \\
& & \; \frac{1}{2} \gamma^2
\epsilon_2\int_0^T\mathrm{d}t\int_0^T\mathrm{d}s
[\tilde{y}(t)y_e(t)+y_e(s)\tilde{y}(s)]F_a^{'}(t-s) + \nonumber \\
& & \; \epsilon_2^2 \frac{1}{2} \gamma^2
\int_0^T\mathrm{d}t\int_0^T\mathrm{d}s
\tilde{y}(t)\tilde{y}(s)F_a^{'}(t-s)=\phi[y_e,y_e]+2\varphi[\tilde{y},
y_e]+\varphi[\tilde{y},\tilde{y}].
\end{eqnarray}

Therefore, from (\ref{pathint}), we have
\begin{eqnarray}
\lefteqn{J(r(T),y(T),r(0),y(0))=} \nonumber \\
& &e^{\frac{i}{\hbar}\tilde{S}[r_e,y_e]}
e^{-\frac{1}{\hbar}\phi[y_e,y_e]}\int_0^0\mathrm{D}\delta y(t)
\mathrm{D}\delta r(t) e^{\frac{i}{\hbar}\tilde{S}[\delta r,\delta y]}
e^{-\frac{2}{\hbar}\varphi[\delta y,y_e]}e^{-\frac{1}{\hbar}
\varphi[\delta y,\delta y]}. \label{superpro2}
\end{eqnarray}

We are now going to show that (\ref{superpro2}) is a function of the
initial and final times only, which is not obvious because of the
functional dependence on $y_e(t)$. In order to do this we discretize
the paths and re-write (\ref{superpro2}) in the form
\cite{feynman2}:
\begin{eqnarray}
\lefteqn{\exp\left\{\frac{i}{\hbar}\tilde{S}[\delta r,\delta
    y]\right\}\approx} \nonumber \\
& &\exp\bigg\{\frac{i}{\hbar}\bigg[\sum_{j=1}^N \epsilon \,m \bigg(\frac
{(\delta r_j-\delta r_{j-1})(\delta y_j-\delta y_{j-1})}{\epsilon^2}
-\omega_0^2\delta r_{j-1}\delta y_{j-1}\bigg) \nonumber \\
& &+\gamma^2\epsilon^2\sum_{j=1}^N\sum_{k=1}^N \delta y_j\delta r_k
F_{a_{(j-k)}}^{''}\bigg]\bigg\},
\end{eqnarray}
where $\delta r_j=\delta r(t_j)$,
$F_{a_{(j-k)}}^{''}=F_a^{''}(t_j-t_k)$,
\begin{eqnarray}
\exp\left\{-\frac{1}{\hbar}\varphi[\delta y,\delta y]\right\}\approx
\exp\left\{-\frac{1}{\hbar}\sum_{j=1}^N
2\gamma^2\epsilon^2\sum_{k=1} ^N \delta y_j\delta y_k
F_{a_{(j-k)}}^{'}\right\}
\end{eqnarray}
and
\begin{eqnarray}
\exp\left\{-\frac{2}{\hbar}\varphi[\delta y,y_e]\right\}\approx
\exp\left\{-\frac{1}{\hbar}\sum_{j=1}^N
4\gamma^2\epsilon^2\sum_{k=1} ^N \delta y_j y_{e_{k}}
F_{a_{(j-k)}}^{'}\right\}.
\end{eqnarray}

Grouping the exponents we obtain
\begin{eqnarray}
\exp{\left\{\frac{i}{\hbar}\tilde{S}[\delta r,\delta y]
-\frac{1}{\hbar}\varphi[\delta y,\delta y]
-\frac{2}{\hbar}\varphi[\delta y,y_e]\right\}} \approx
\exp{\left\{-\frac{i}{2}U^T M U -A^T U \right\}},
\end{eqnarray}
with
\begin{eqnarray}
U^T\equiv(\delta r_1 \ldots \delta r_N \, \delta y_1 \ldots \delta y_N)
\qquad M\equiv \left( \begin{array}{cc}
 0 & p \\
 p & r
\end{array} \right),
\end{eqnarray}
and where $p$ and $r$ are $N$x$N$ matrices and $A^T=(0 \; a)$ and
$a$ are $N$-dimensional vectors. To solve the path integral we need
to integrate this exponent over $\mathrm{d}U=\mathrm{d}\delta r_1
\ldots \mathrm{d}\delta r_N \, \mathrm{d}\delta y_1 \ldots \delta
y_N$. The result is \cite{swanson}
\begin{eqnarray}
\frac{1}{(\mathrm{det}\, M)^{1/2}}\exp\left[-\frac{1}{4}A^T M^{-1} A \right].
\end{eqnarray}

Because $M$ has a zero upper left block, its inverse has a zero
lower right block and, therefore, $A^T M^{-1}A = 0$. Since all the
dependence on the initial and final positions is contained in $A$,
(\ref{superpro2}) is indeed a function only of the initial and final
times. Therefore we may write the superpropagator as
\begin{eqnarray}
J(r(T),y(T),r(0),y(0))=G(T,0)\exp\left\{\frac{i}{\hbar}\tilde{S}[r_e,y_e]
\right\}\exp\left\{-\frac{1}{\hbar}\phi[y_e,y_e]\right\},
\end{eqnarray}
and $G(T,0)$ can be calculated by imposing the normalization of the
reduced density operator.

%%%%%%%%%%%%%%%%%%%%%%%%%%%%%%%%%%%%%%%%%%%%%%%%%%%%%%%%%%%%%%
%%%%%%%%%%%%%%%%%%%%%%%%%%%%%%%%%%%%%%%%%%%%%%%%%%%%%%%%%%%%%%
\section{Solution of the Equations of Motion.}

Taking the Laplace transform of (\ref{eq22}), we get (with
$F^{''}_a(t)\approx\phi_{xx}(t)/2$)
\begin{eqnarray}
\left[(s^2+\Omega_0^2)-\frac{\gamma^2}{m}\tilde{\phi}_{xx}(s)\right]
\tilde{r}_e(s)=sr(0)+\dot{r}(0).\label{a2-1}
\end{eqnarray}
where $\tilde{f}(s)=\mathcal{L}\{f(t)\}$ is the Laplace transform of
$f(t)$. Using
\begin{eqnarray}
\phi_{xx}(t)=\frac{2}{E_c(0)}\langle
p_x(0)x(t)\rangle_e=A\,e^{-\alpha|t|}\sin{(\omega t)}, \label{a2-3}
\end{eqnarray}
(\ref{a2-1}) becomes
\begin{eqnarray}
\tilde{r}_e(s)=\frac{s[(s+\alpha)^2+\omega^2]r(0)+[(s+\alpha)^2+\omega^2]
\dot{r}(0)}{\{(s^2+\Omega_0^2)[(s+\alpha)^2+\omega^2]-
\frac{\gamma^2}{m}A\omega\}}.\label{a2-4}
\end{eqnarray}

The Heaviside's theorem establishes that if $P(s)$ and $Q(s)$ are
polynomials such that the order of $P(s)$ is smaller than the order
of $Q(s)$, then
\begin{eqnarray}
\mathcal{L}^{-1}\left[\frac{P(s)}{Q(s)}\right]=\sum_{i=1}^{n}
\frac{P(s_i)}{Q'(s_i)} e^{s_it},\label{a2-5}
\end{eqnarray}
where $s_i$ are the roots of $Q(s)=0$ and $Q'(s)$ is the
$s$-derivative of $Q(s)$. Therefore we need the roots of
\begin{eqnarray}
\left[x^2+\left(\frac{\omega_0}{\alpha}\right)^2\right]\left[(x+1)^2+
\left(\frac{\omega}{\alpha}\right)
^2\right]-\frac{\gamma^2}{m}\frac{A\omega}{\alpha^4}=0,
\label{a2-7}
\end{eqnarray}
where $x=s/\alpha$ and $\Omega_0\approx \omega_0$. From Section II
we have
\begin{eqnarray}
\left(\frac{\omega_0}{\alpha}\right)^2\approx 1.6\times 10^{-2}\qquad
\left(\frac{\omega}{\alpha}\right)^2\approx 25 \qquad
\frac{\gamma^2}{m}\frac{A\omega}{\alpha^4}\approx 3\times 10^{-2}.\label{a2-8}
\end{eqnarray}
and the roots of (\ref{a2-7}) are
\begin{eqnarray}
x_1=-1.00-i 5.00 &\qquad& x_2=-1.00+i5.00
\nonumber \\
x_3=-4\times10^{-5}-i0.12 &\qquad& x_4=-4\times10^{-5}+i0.12,\label{a2-9}
\end{eqnarray}
Multiplying these roots by $\alpha$, we get
\begin{eqnarray}
s_1\approx-\alpha-i \omega &\qquad& s_2\approx-\alpha+i\omega
\nonumber \\
s_3\approx-\Lambda-i\omega_0 &\qquad& s_4\approx
-\Lambda+i\omega_0.\label{a2-10}
\end{eqnarray}
The same procedure is applied to (\ref{eq23}). The Laplace transform
of (\ref{eq23}) is written as
\begin{eqnarray}
\tilde{y}_e(s)=\frac{s[(s+\alpha)^2+\omega^2]y(0)+[(s+\alpha)^2+\omega^2]
\dot{y}(0)}{\{(s^2+\Omega_0^2)[(s+\alpha)^2+\omega^2]+
\frac{\gamma^2}{m}A\omega\}}\label{a2-11}
\end{eqnarray}
and the roots are
\begin{eqnarray}
s_1\approx-\alpha-i \omega &\qquad& s_2\approx-\alpha+i\omega
\nonumber \\
s_3\approx \Lambda-i\omega_0 &\qquad& s_4\approx
\Lambda+i\omega_0.\label{a2-12}
\end{eqnarray}

Since we are interested on time scales such that $t\sim 1/\omega_0$,
$s_1$ and $s_2$ are transient solutions and only $s_3$ and $s_4$ are
important. Therefore, turning to the equations (\ref{eq-3}) and
(\ref{eq-4}) and considering times on the scale $t\sim
1/\omega_0$, we see that those equations can be rewritten
approximately as
\begin{eqnarray}
\ddot{r}_e(t)+2\Lambda\dot{r}_e(t)+\Omega_0^2r_e(t)=0,\label{a2-13} \\
\ddot{y}_e(t)-2\Lambda\dot{y}_e(t)+\chi_0^2y_e(t)=0,\label{a2-14}
\end{eqnarray}
where terms proportional to $F(t)$ were disregarded (since they go to
zero for $t\sim 1/\omega_0$) and the convolutions terms were
approximated in the following way
\begin{eqnarray}
\int_0^t\mathrm{d}s\,F(t-s)\dot{r}_e(s)\approx\dot{r}_e(t)
\lim_{t\to\infty}\int_0^{t}\mathrm{d}s
F(t-s).\label{a2-15}
\end{eqnarray}
Thus, $\Lambda$ is given by
\begin{eqnarray}
\Lambda=\frac{\gamma^2}{2m}\lim_{t\to\infty}\int_0^{t}\mathrm{d}s
F(t-s).\label{a2-16}
\end{eqnarray}

Indeed, applying the Laplace transform in (\ref{a2-13}) and
(\ref{a2-14}), we get the roots
\begin{eqnarray}
s_1=-\Lambda-i\omega_0 &\qquad& s_2=-\Lambda+i\omega_0,\label{a2-17}
\end{eqnarray}
for $r_e(t)$ and
\begin{eqnarray}
s_1=\Lambda-i\omega_0 &\qquad& s_2=\Lambda+i\omega_0,\label{a2-18}
\end{eqnarray}
for $y_e(t)$ since $\Omega_0^2,\chi_0^2\gg\Lambda$ and
$\Omega_0^2\approx\chi_0^2\approx\omega_0^2$. Comparing
(\ref{a2-17}) and (\ref{a2-18}) with (\ref{a2-10}) and
(\ref{a2-12}), we conclude that the equations (\ref{a2-13}) and
(\ref{a2-14}) give a good description of the behavior given by
(\ref{eq-3}) and (\ref{eq-4}) for $t\sim 1/\omega_0$.

\end{appendix}

%%%%%%%%%%%%%%%%%%%%%%%%%%%%%%%%%%%%%%%%%%%%%%%%%%%%%%%%%%%%%%
%%%%%%%%%%%%%%%%%%%%%%%%%%%%%%%%%%%%%%%%%%%%%%%%%%%%%%%%%%%%%%%%%%

\centerline{\bf Acknowledgements} \noindent This paper was partly
supported by the Brazilian agencies {\bf FAPESP}, under contracts
number 02/04377-7 and 03/12097-7, and {\bf CNPq}. Especial thanks to
S.M.P.

%%%%%%%%%%%%%%%%%%%%%%%%%%%%%%%%%%%%%%%%%%%%%%%%%%%%%%%%%%%%%%%%%
%%%%%%%%%%%%%%%%%%%%%%%%%%%%%%%%%%%%%%%%%%%%%%%%%%%%%%%%%%%%%%%%%

\newpage
%%%%%%%%%%%%%%%%%%%%%%%%%%%%%%%%%%%%%%%%%%%%%%%%%%%%%%%%%%%%%%%%
%%%%%%%%%%%%%%% FIGURES %%%%%%%%%%%%%%%%%%%%%%%%%%%%%%%%%%%%%%%%
%%%%%%%%%%%%%%%%%%%%%%%%%%%%%%%%%%%%%%%%%%%%%%%%%%%%%%%%%%%%%%%%
\begin{figure}
\centering
\includegraphics[clip=true,width=6cm,angle=0]{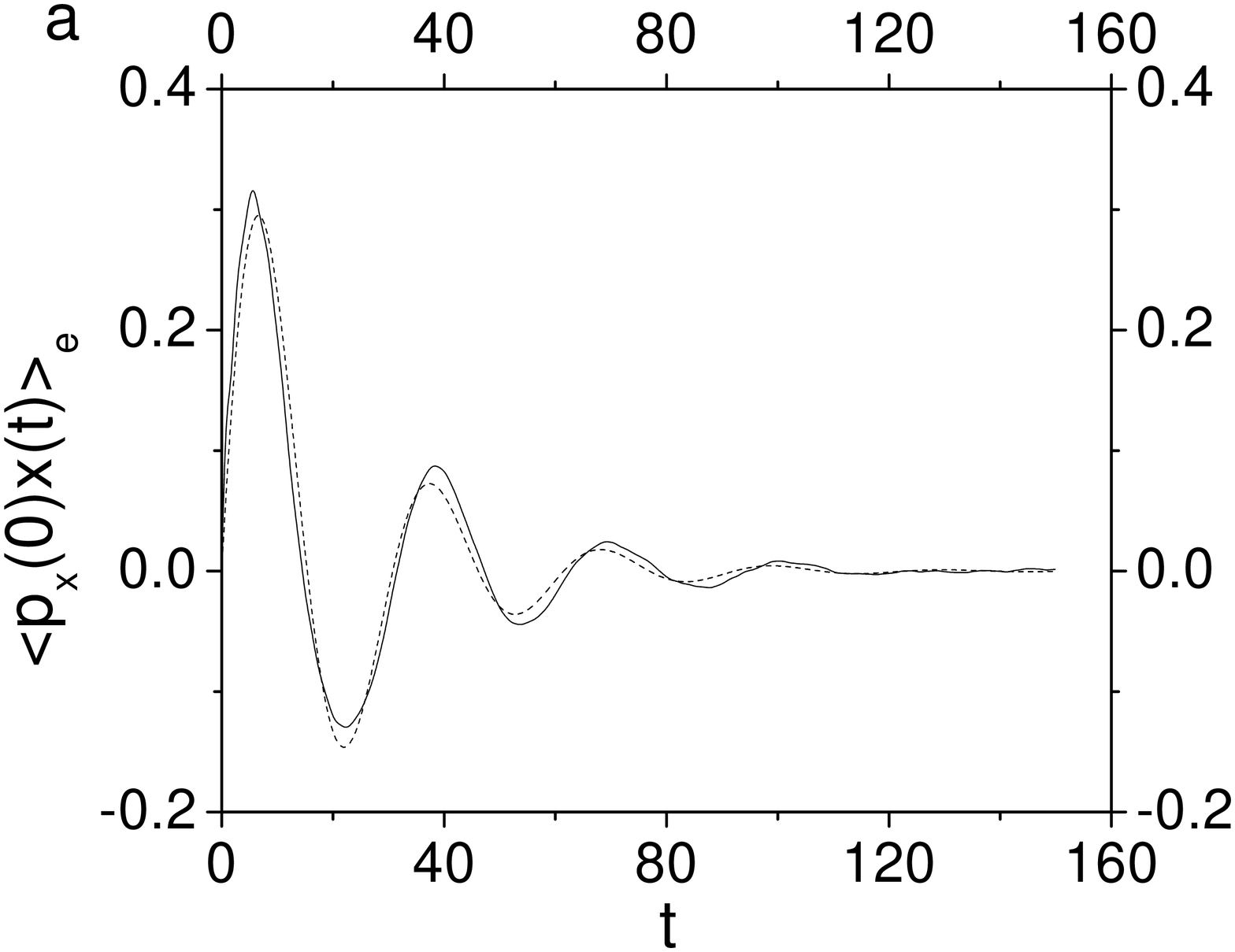}
\includegraphics[clip=true,width=6cm,angle=0]{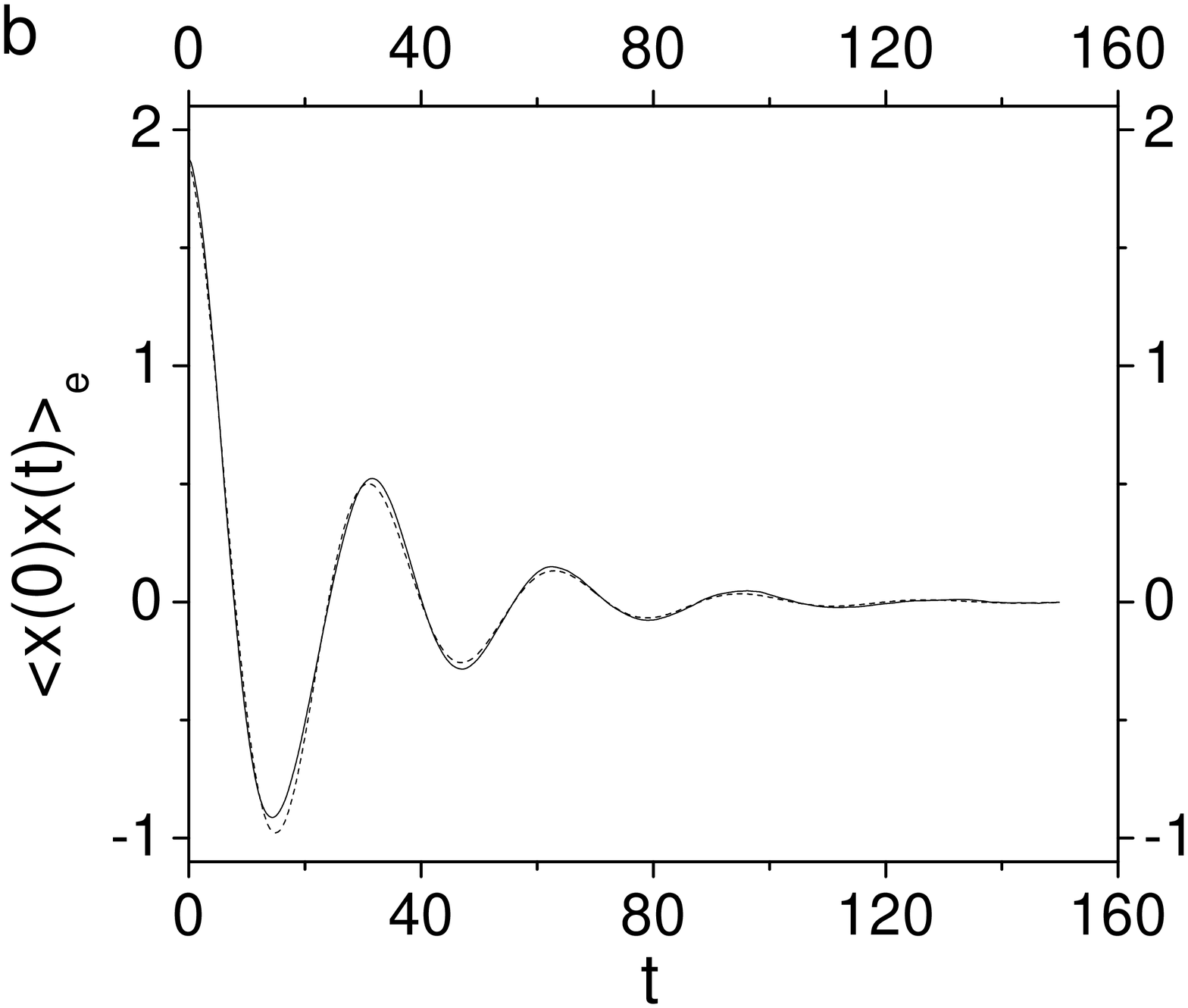}
\label{fig2} \caption{Correlation functions for the NS for
$E_c=0.38$: (a) $\langle p_x (0)x(t)\rangle_e$; (b) $\langle
x(0)x(t)\rangle_e$. The full line shows the numerical results and
the dashed line shows the fitting. The averages were computed using
35000 initial conditions.}
\end{figure}

\begin{figure}
\centering
\includegraphics[clip=true,width=5cm,angle=0]{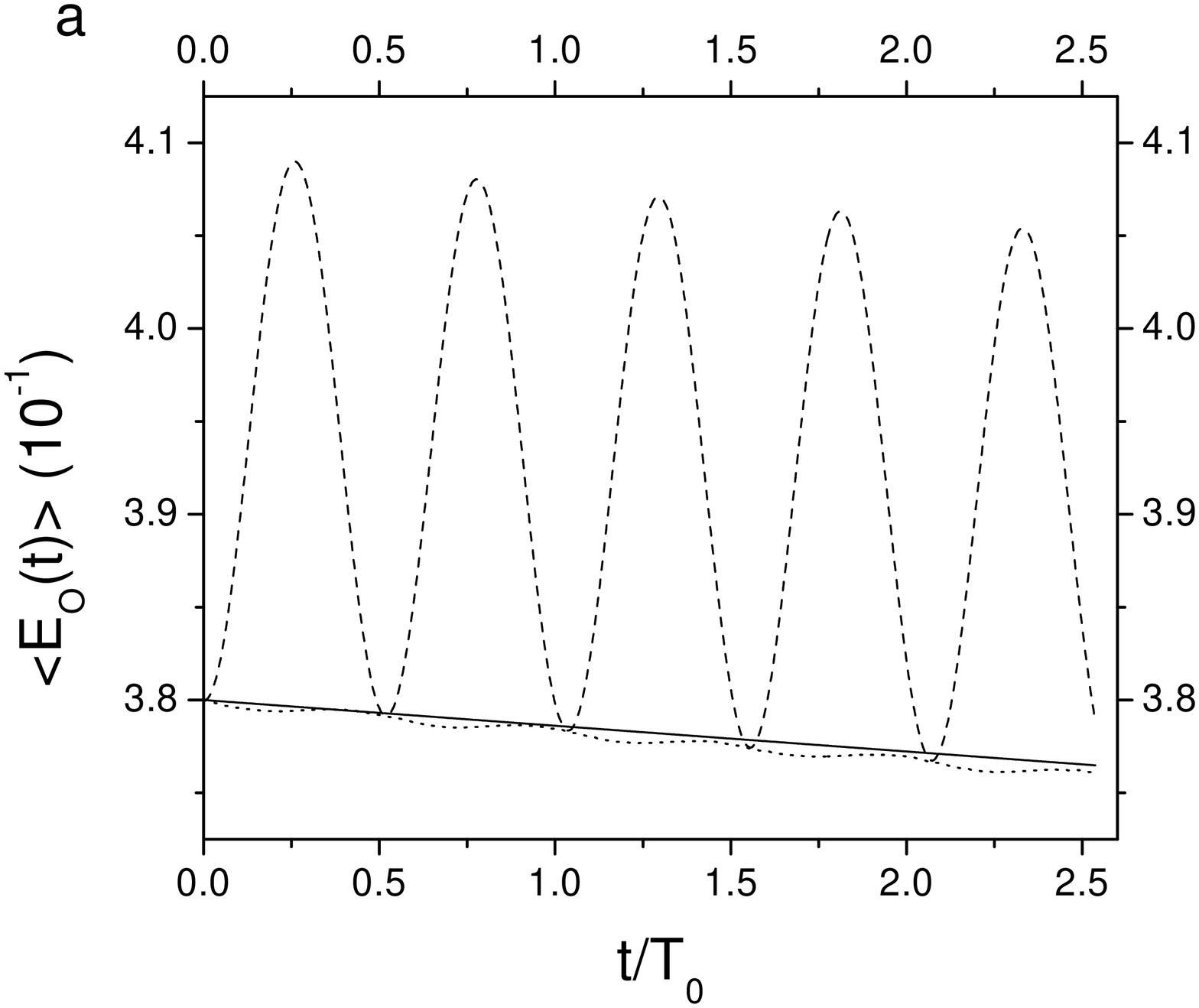}
\includegraphics[clip=true,width=5cm,angle=0]{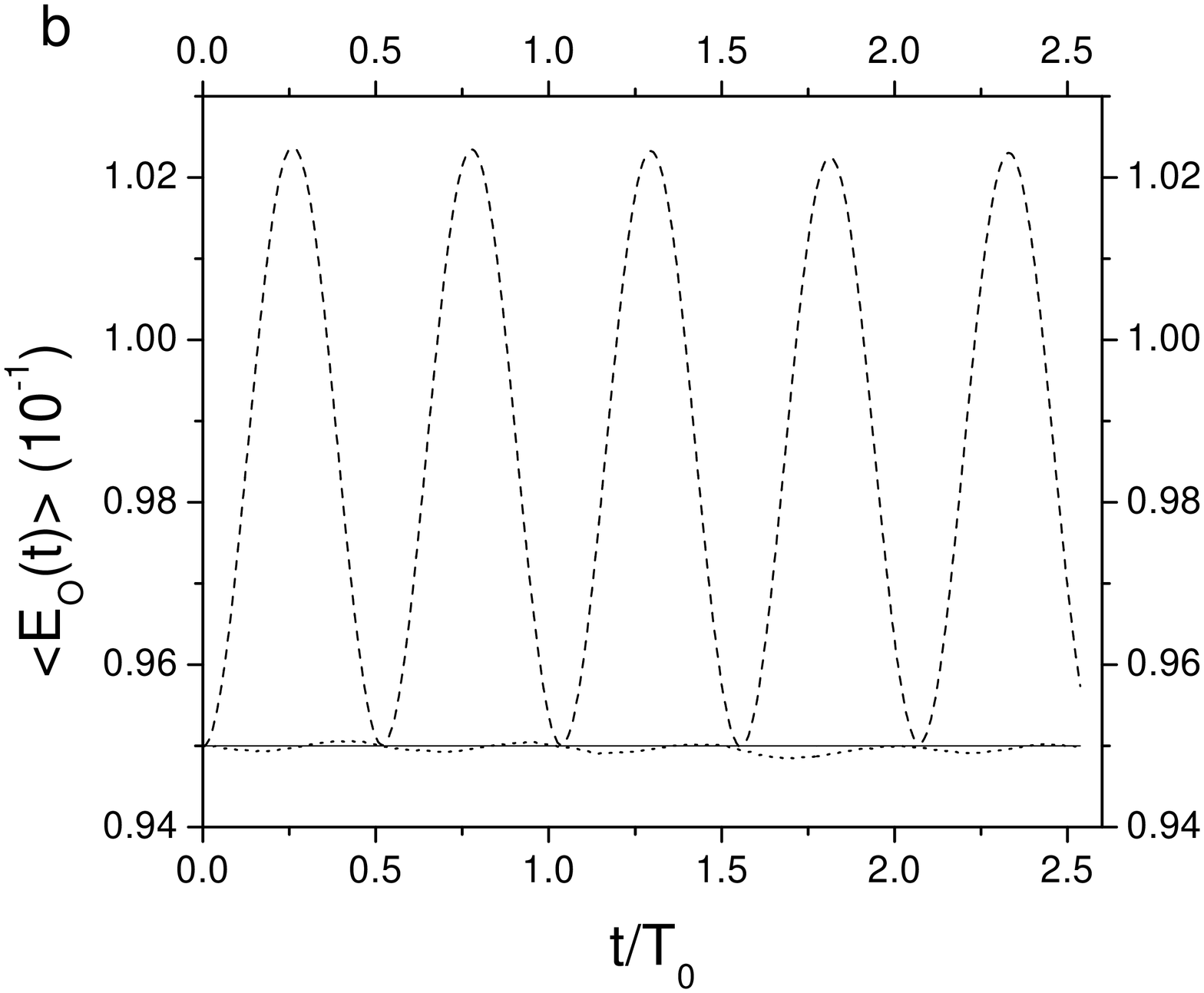}
\includegraphics[clip=true,width=5cm,angle=0]{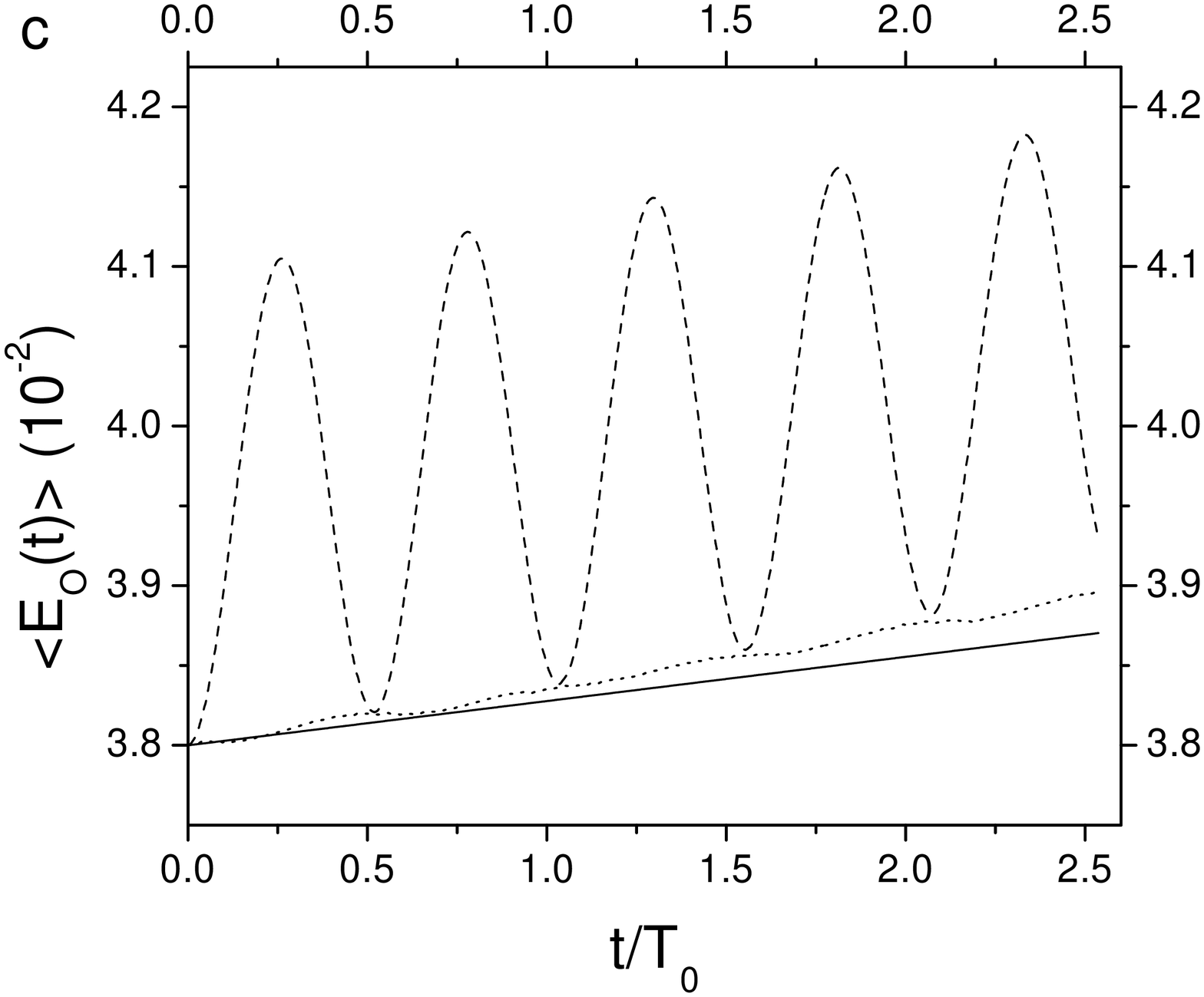}
\label{fig3} \caption{Average oscillator energy at short times
  with the NS as chaotic system. $T_0=1/\omega_0$. The dashed line
  shows $\langle
  E_o(t)\rangle$ and the doted line shows $\langle
  E_{or}(t)\rangle$, both obtained numerically. The full line
  shows Eq.(\ref{eq13}) without $f(t)$. (a)
  $E_o(0)/E_c(0)=1.0$, (b) $E_o(0)/E_c(0)=0.25$ and (c)
  $E_o(0)/E_c(0)=0.1$. The oscillator's parameters, coupling
  constant and number of initial conditions are the same as in Fig.2.}
\end{figure}

\begin{figure}
\centering
\includegraphics[clip=true,width=8.7cm,angle=0]{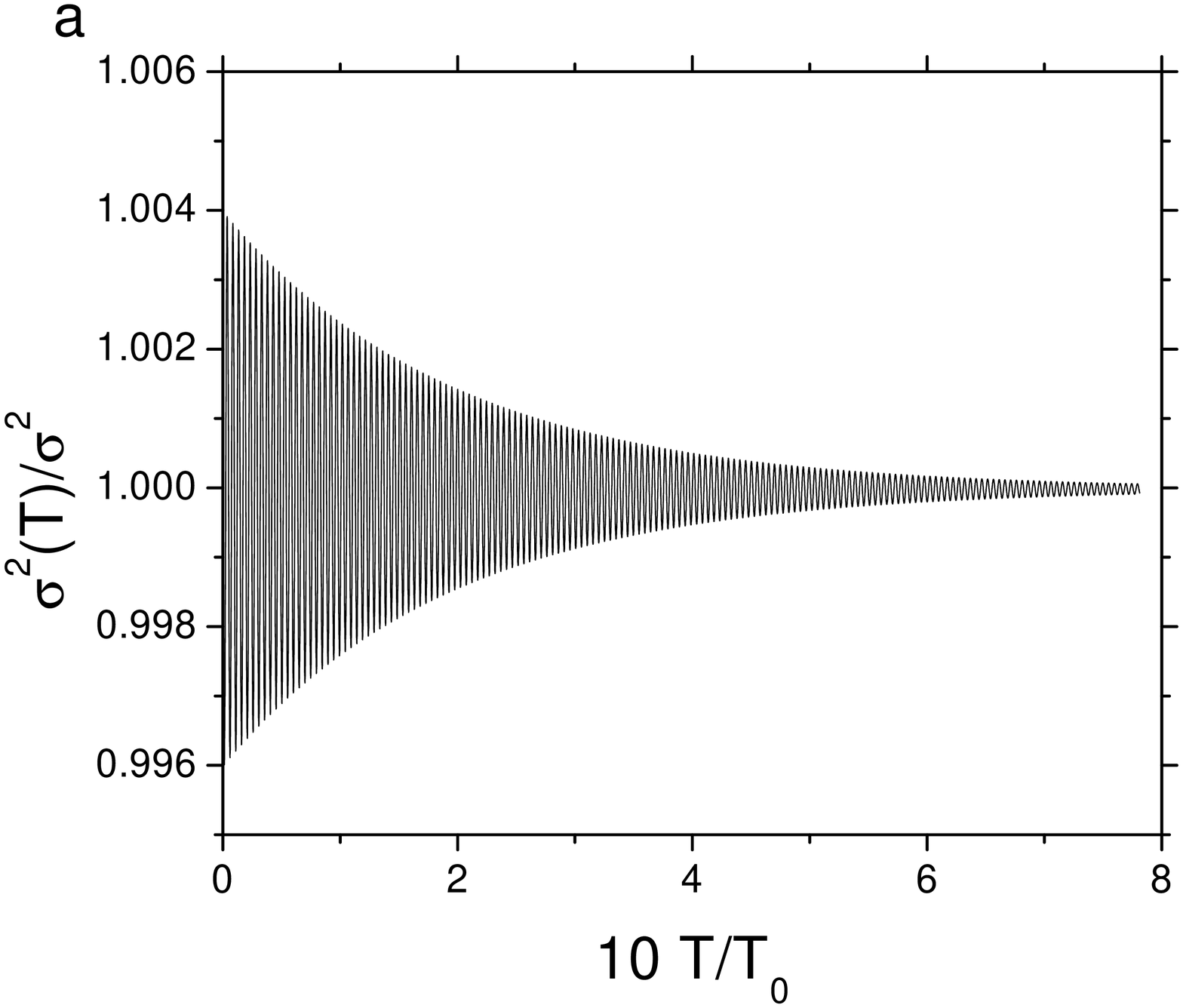}
\includegraphics[clip=true,width=8.7cm,angle=0]{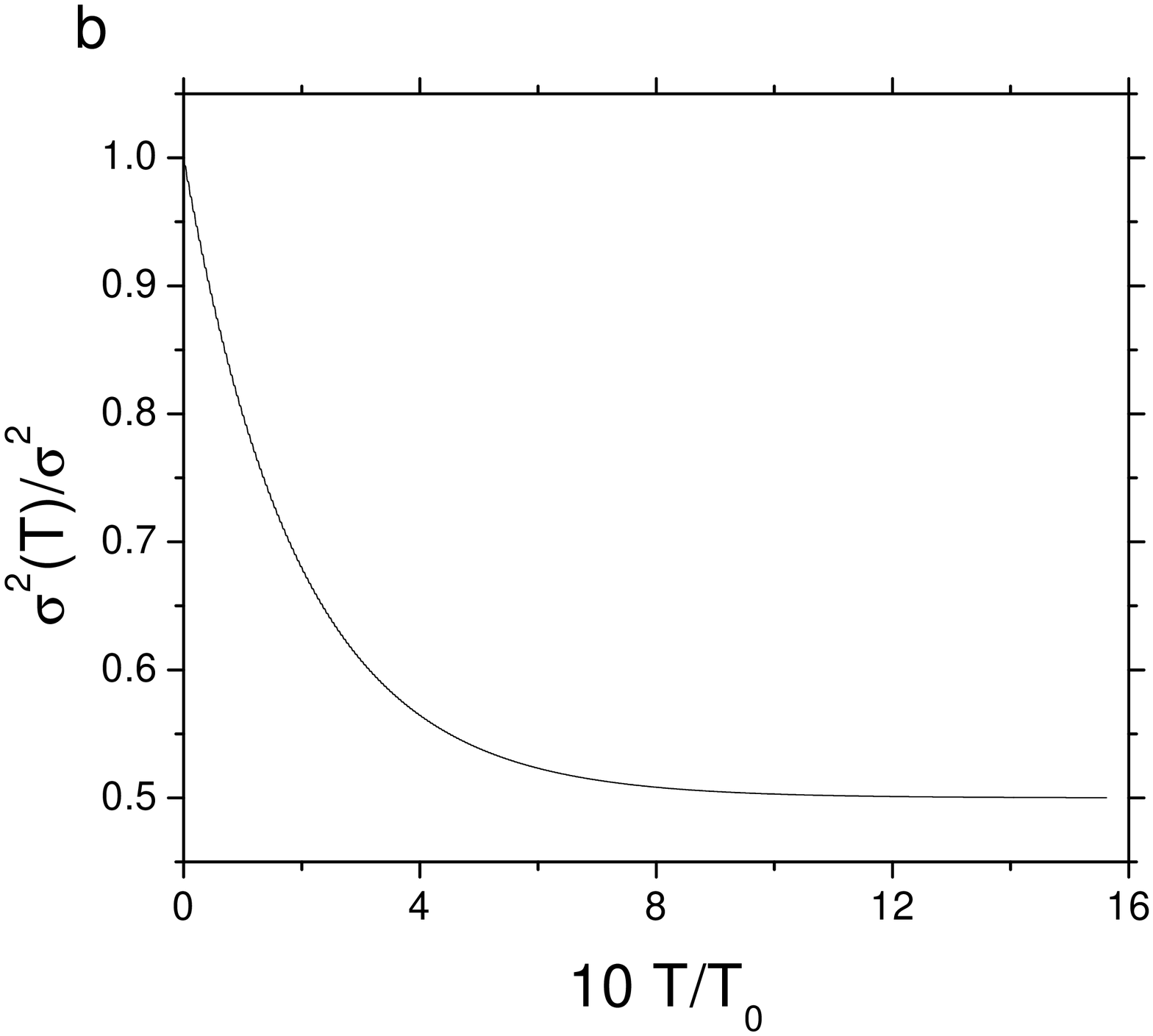}
\includegraphics[clip=true,width=8.7cm,angle=0]{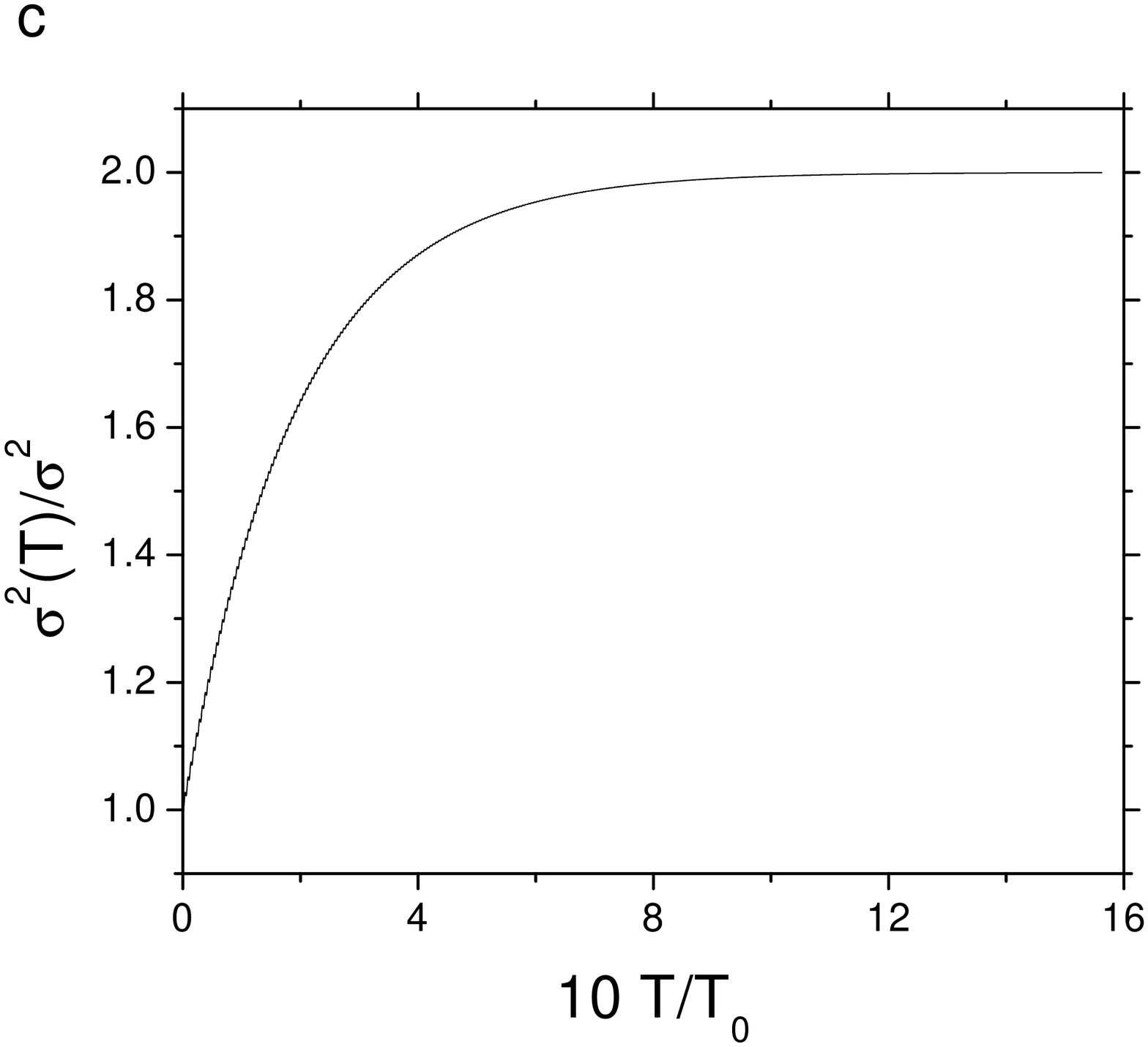}
\label{fig4} \caption{Squared width of wave packet
$\sigma^2(T)/\sigma^2$ as given by Eq.(\ref{eq-35}) with
$T_0=1/\omega_0$.
  (a) $\Gamma=1.0$, (b) $\Gamma=0.5$ and (c) $\Gamma=2.0$.}
\end{figure}

\begin{figure}
\centering
\includegraphics[clip=true,width=9cm,angle=0]{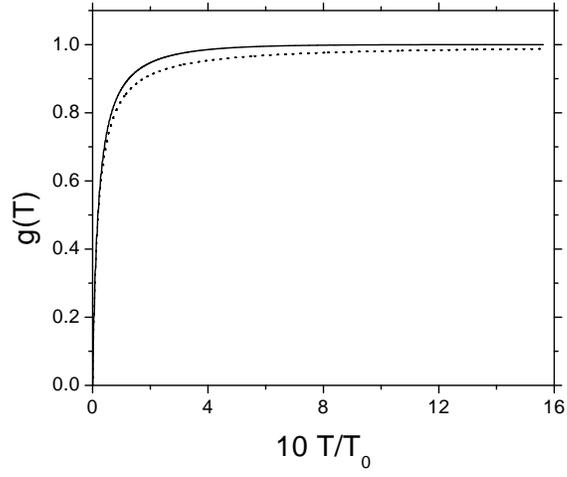}
\label{fig5} \caption{The full line shows $g(T)$ as given by
Eq.(\ref{eq-49}) for $\Gamma=10.0$. $T_0=1/\omega_0$. The dotted
line shows $g(T)$ as given by Eq.(\ref{eq-52}) for $\Gamma=10.0$.}
\end{figure}

\end{document}